\documentclass[fleqn,usenatbib]{mnras}
\usepackage{newtxtext,newtxmath}
\usepackage[T1]{fontenc}
\usepackage{graphicx}
\usepackage{amsmath}
\usepackage{bm}
\usepackage{tikz}
\usetikzlibrary{positioning,calc}
\usepackage{color}
\usepackage{float}

\title[Cosmological Variation of $\mu$]{Cosmological Variation of Proton-to-Electron Mass Ratio in a Chameleon-Brans-Dicke Theory}

\author[Malavika K and S. Chakrabarti]{
Malavika K$^{1}$\thanks{E-mail: malavika.k2024@vitstudent.ac.in}
and
Soumya Chakrabarti$^{1}$\thanks{E-mail: soumya.chakrabarti@vit.ac.in}
\\
$^{1}$School of Advanced Sciences, Vellore Institute of Technology,\\
Tiruvalam Rd, Katpadi, Vellore, Tamil Nadu 632014, India
}

\date{Accepted XXX. Received YYY; in original form ZZZ}

\pubyear{\the\year{}}

\begin{document}

\label{firstpage}
\pagerange{\pageref{firstpage}--\pageref{lastpage}}

\maketitle

\begin{abstract}
We investigate the cosmological variation of the proton-to-electron mass ratio, $\mu=m_p/m_e$, within a generalized Brans-Dicke framework with two scalar fields : a geometric scalar field governing the effective gravitational coupling and a matter field whose effective potential has a Higgs-like symmetry-breaking structure. We consider a cosmological background described by a Pad\'e-deformed $\Lambda$CDM and constrain it using a combination of late-time cosmological observational data-sets. We use the resulting expansion history to reconstruct the scalar-field and the matter-sector vacuum expectation value. We calculate the induced variation of $\mu$ and compare it with observational constraints from quasar absorption spectra. We find that a standard Brans-Dicke model with a minimally coupled Higgs-like field produces an unphysical variation of $\mu$. Only after introducing a chameleon-like coupling between the Brans-Dicke scalar and the baryonic matter, we find that the resulting variation is well-suppressed and satisfies the observational bounds.
\end{abstract}


\section{Introduction}
The experimental discovery of the Higgs boson has established scalar fields as fundamental ingredients of the Standard Model of particle physics \citep{Aaboud_2018, CMS}. Scalar fields lead to additional scalar degrees of freedom and these also play an important role in cosmology and modified theories of gravity. In cosmology, scalar fields have been invoked in a variety of contexts, for instance in theories of inflation \citep{Guth_2005, PhysRevLett.62.376}, quintessence \citep{Tsujikawa_2013} and dilaton fields \citep{Grumiller_2002, PhysRevD.82.076009, PhysRevD.106.063540}. Moreover, within the conventional approach, the late-time acceleration of the universe is usually attributed to an additional component with effective negative pressure. This is commonly referred to as Dark Energy (DE) \citep{Riess_2001, Betoule_2014, PhysRevLett.86.6, Upadhye_2005} and also most popularly modelled by scalar-fields \citep{PhysRevLett.82.896, COPELAND_2006, PhysRevD.63.103510, Caldwell:2003vq, PhysRevD.66.021301, 10.1111/j.1365-2966.2009.15015.x}. At the same time, scalar fields constitute a central ingredient of modified theories of gravity, where non-minimal couplings can generate deviations from General Relativity (GR) while allowing local screening in high-density environments. Chameleon \citep{Brax:2004px}, Symmetron \citep{PhysRevD.84.103521}, and Galileon \citep{PhysRevD.80.024037} models are prominent examples in this category. Apart from standard applications, categorically different approaches such as the holographic dark energy \citep{Li:2004rb} or the emergent cosmological scenarios \citep{PhysRevD.104.063521}, further illustrate the importance of additional dynamical degrees of freedom in cosmology.  \\

Brans-Dicke (BD) theory \citep{BransDicke1961} is an interesting prototype : a natural scalar extension of GR in which the effective gravitational coupling is not a constant, but rather a scalar field $G_{\mathrm{eff}} \propto 1/\phi$. A dimensionless parameter $\omega$ is introduced by hand to control the deviation from GR. The GR limit is formally recovered at $\omega\to\infty$ \citep{PhysRevD.56.1334, PhysRevD.80.124040}. Observational constraints require any present-day variation of the gravitational coupling to be very small \citep{Uzan2011}, but a mild evolution of $G_{\mathrm{eff}}$ in the cosmological past remains a theoretically interesting possibility. Indeed there are consideration that effective gravitational coupling and standard model parameters evolve simultaneously, as a function of an underlying energy scale. This receives motivation from the renormalization-group (RG) description of quantum field theories \citep{Dou_1998, Elizalde_1995, PhysRevD.85.104016, PhysRevLett.133.021604, BirrellDavies1982}. There are quite a few studies where the evolution of an effective gravitational coupling is considered to be a function of Hubble scale \citep{Sol__2013, PhysRevD.110.024055}. Motivated by these, we consider a generalized BD framework containing two scalar fields : the geometric BD scalar $\phi$ and a matter-sector scalar $\psi$. The second scalar field is minimally coupled and has an effective Higgs-like structure, allowing a cosmological variation of vacuum expectation value (VEV). We parameterize the variation of gravitational coupling through a logarithmic form \citep{BirrellDavies1982, Shapiro_2005, deCruzPerez:2023wzd}
\begin{equation}
G_{\mathrm{eff}}(H) = \frac{G_0}{1+2\nu_G\ln(H/H_0)},
\end{equation}
popular in RG-improved theories of gravity as well as in running-vacuum scenarios \citep{Fritzsch_2017, Basilakos_2009, Shapiro_2009, Sola:2016vis, Fritzsch_2015, Fritzsch_2012}. \\

We deliberately avoid using the original, single scalar BD framework, mainly because of the constraints on $\omega$ coming from local gravitational tests. There are several generalized scalar-tensor frameworks which allow $\omega$ to vary as a function of the BD field itself \citep{Nordtvedt:1970uv, PhysRevD.1.3209, 1978ApJ2195B, PhysRevD.30.728, PhysRevD.73.104022, PhysRevD.78.043512}. However, we find that a particularly interesting possibility is to introduce a non-minimal coupling between the BD scalar field and ordinary, baryonic matter distribution \citep{PhysRevLett.93.171104, 10.1111, PhysRevD.73.083509}. This coupling allows the scalar interaction to vary depending on the background density. In effect, the BD scalar behaves like a chameleon and a screening mechanism is brought in to effect \citep{PhysRevD.75.104016, HarkoLobo2010, Khoury_2004, Mota_2011, PhysRevD.86.127503}. It has also been proved recently that such a Chameleon-Brans-Dicke theory can relax the constraints on the BD parameter $\omega$ coming from solar system experiments. Therefore, we consider a Chameleon-Brans-Dicke-Higgs framework in which the geometric scalar is non-minimally coupled to the matter Lagrangian, while the additional Higgs-like scalar $\psi$ remains minimally coupled.  \\

\begin{figure}
\centering
\begin{tikzpicture}[
    node distance=5mm,
    box/.style={
        draw,
        rounded corners=2pt,
        align=center,
        minimum width=3.05cm,
        minimum height=0.65cm,
        inner sep=4pt
    },
    arrow/.style={->, thick},
    every node/.style={font=\scriptsize}
]


\node[box] (Hz)
    {Cosmological background\\$H(z)$};

\node[box, below=of Hz] (Geff)
    {Effective gravitational coupling\\
     $G_{\mathrm{eff}}(H)$};

\node[box, below=of Geff] (phi)
    {Geometric scalar\\$\phi(z)$};


\node[box, below=8mm of phi] (RG)
    {Renormalization-group\\motivation};


\node[box, above right=9mm and 12mm of phi] (cham)
    {Chameleon interaction\\
     $f(\phi)$};


\node[box, right=12mm of phi] (psi)
    {Matter-sector scalar\\$\psi(z)$};


\node[align=center, font=\scriptsize]
    (param)
    at ($(phi)!0.65!(cham)!0.30!(psi)$)
    {Parametric\\reconstruction};


\node[box, below=of psi] (potential)
    {Higgs-like potential\\$V(\psi,z)$};

\node[box, below=of potential] (vev)
    {Vacuum expectation value\\$\nu(z)$};

\node[box, below=of vev] (mu)
    {Proton-to-electron mass ratio\\
     $\mu(z)=m_p/m_e$};

\node[box, below=of mu] (obs)
    {Quasar Absorption Constraints};


\draw[arrow] (Hz) -- (Geff);
\draw[arrow] (Geff) -- (phi);


\draw[arrow] (RG) -- (phi);


\draw[arrow] (phi.east) -- (cham.south west);

\draw[arrow] (cham.south) -- (psi.north);

\draw[arrow] (psi.west) -- (phi.east);


\draw[arrow] (psi) -- (potential);
\draw[arrow] (potential) -- (vev);
\draw[arrow] (vev) -- (mu);
\draw[arrow] (mu) -- (obs);


\end{tikzpicture}

\caption{
Schematic diagram of the reconstruction adopted in this work.
The cosmological expansion history determines the effective gravitational
coupling $G_{\mathrm{eff}}(H)$ and the geometric scalar field $\phi(z)$,
with the latter motivated by the renormalization-group description of
the running gravitational coupling. In the interacting extension, the
geometric scalar is coupled to the matter sector through the
chameleon-type interaction $f(\phi)$. The matter-sector scalar
$\psi(z)$ is retained as a minimally coupled field. The closed loop
formed by $\phi(z)$, $f(\phi)$, and $\psi(z)$ represents the parametric
reconstruction of the interacting scalar sector. The reconstructed
matter field subsequently determines the Higgs-like potential, its
vacuum expectation value $\nu(z)$, and the proton-to-electron mass ratio
$\mu(z)$, which is confronted with quasar absorption constraints.
}
\label{fig:reconstruction_flow}
\end{figure}
Our aim is to reconstruct the matter-sector scalar field from the generalized BD field equations, based on a prescribed cosmic expansion history. The $\Lambda$CDM model provides a successful and widely adopted framework for describing the late-time expansion of the Universe, although there are several conceptual and theoretical issues related to the model. Several phenomenological parametrizations have been proposed to describe small departures from the standard $\Lambda$CDM, including the Chevallier-Polarski-Linder (CPL) parametrization \citep{Chevallier:2000qy, PhysRevLett.90.091301, Linden:2008mf}, alternative two-parameter forms \citep{Jassal:2004ej, PhysRevD.72.103503, Barboza:2008rh, PhysRevD.77.123525, Wetterich:2004pv} and Quintom scenarios \citep{Feng:2004ad, Guo:2004fq}. Rather than adopting one of these, we employ a simple rational Pad\'e-deformation of the $\Lambda$CDM Hubble function. The Pad\'e coefficients allow a controlled deviation from the standard $\Lambda$CDM behavior, while the present day cosmographic behavior is recovered when the deformation parameters vanish. We constrain the Pad\'e parameters through a Bayesian analysis using a combination of cosmological observations and use the resulting expansion history as the background dynamics. We solve for the matter-like field whose self-interaction is Higgs-like with a symmetry-breaking structure and a dynamically evolving effective mass term \citep{Wetterich1988, FerreiraJoyce1997}. We derive the effective mass term and determine the redshift dependence of the vacuum expectation value (VEV). Since the VEV sets the relevant fermion mass scale in the model, its cosmological evolution leads to a variation of particle masses and, in particular, to a variation of the proton-to-electron mass ratio $\mu=m_p/m_e$. We compare the reconstructed variation of $\mu$ with observational constraints obtained from spectroscopic measurements of quasar absorption systems. For a minimally coupled generalized BD framework with no chameleon interaction, we find that the resulting variation of $\mu$ is unphysical and incompatible with the Quasar spectroscopic bounds. However, once the chameleon coupling is switched on, the BD scalar field helps suppress the evolution of the matter-field VEV. As a result the variation of $\mu$ remains well within the observationally allowed variation, in the low-redshift regime. Finally, we examine the simultaneous evolution of the effective gravitational coupling $G_{\mathrm{eff}}$ and the proton-to-electron mass ratio $\mu$, providing a unified view of the cosmological evolution of the gravitational and particle-mass sectors. This flow of the investigation is summarized in Chart \ref{fig:reconstruction_flow}.  \\

In the next section, we discuss the origin of Renormalization Group (RG) motivated running of the effective gravitational coupling. In Section $3$, we discuss the mechanism responsible for a variation of Higgs vacuum expectation value. In Section $4$, we introduce the Pad\'e-deformed $\Lambda$CDM expansion history. The reconstruction of proton-to-electron mass ratio $\mu$, based on all of these mechanisms, is discussed in detail in Section $5$. We make some concluding remarks in Section $6$. 

\section{Geometric Scalar Sector : RG-motivated variation of Gravitational Coupling}
\label{sec:runningG}

One of the central ingredients of this work is the assumption that the effective gravitational coupling evolves with the cosmological energy
scale. In a scalar-tensor framework such as Brans-Dicke theory, the effective Newtonian coupling is controlled by the geometric scalar,
$G_{\mathrm{eff}}\propto\phi^{-1}$. It is natural to investigate whether the cosmological evolution of the scalar can be related to the evolution of a characteristic energy scale associated with the background expansion. The Hubble parameter $H$ provides such a scale and we parameterize the effective gravitational coupling as \citep{BirrellDavies1982, Shapiro_2005, deCruzPerez:2023wzd}
\begin{equation}
G_{\mathrm{eff}}(H) = \frac{G_0}{1+2\nu_G\ln(H/H_0)},
\label{eq:runningG}
\end{equation}
where $G_0$ and $H_0$ denote the present-day values. $\nu_G$ is a dimensionless parameter controlling the strength of the variation. The normalization is such that
\begin{equation}
G_{\mathrm{eff}}(H_0)=G_0.
\end{equation}
For $\lvert\nu_G\ln(H/H_0)\rvert\ll1$, Eq. \eqref{eq:runningG} reduces to
\begin{equation}
G_{\mathrm{eff}}(H) \simeq G_0\left[1-2\nu_G\ln\left(\frac{H}{H_0}\right)\right],
\end{equation}
showing explicitly that $\nu_G$ measures the leading logarithmic departure from a constant gravitational coupling ($\nu_G \simeq 10^{-3}$). The logarithmic variation is motivated by the scale dependence of effective couplings in quantum field theory-inspired and by RG-improved descriptions of gravitational interactions \citep{BirrellDavies1982, Shapiro_2005, deCruzPerez:2023wzd}. Some noteworthy examples in this category are running-vacuum models and RG-inspired cosmological frameworks \citep{Fritzsch_2017, Basilakos_2009, Shapiro_2009, Sola:2016vis, Fritzsch_2015, Fritzsch_2012}. In the present work, we use Eq.~\eqref{eq:runningG} as a phenomenological definition. Using the definition of Brans-Dicke scalar field $G_{\mathrm{eff}}\propto\phi^{-1}$, the running in Eq.~\eqref{eq:runningG} can equivalently be expressed, up to the normalization convention for $\phi$, as
\begin{equation}
\frac{\phi(H)}{\phi_0} = 1+2\nu_G\ln\left(\frac{H}{H_0}\right).
\label{eq:phiH}
\end{equation}
Thus, once an expansion history $H(z)$ is specified, one simply has
\begin{equation}
\phi(z) = \phi_0 \left[ 1+2\nu_G\ln\left(\frac{H(z)}{H_0}\right)\right].
\label{eq:phiz}
\end{equation}
This relation provides the first step of the reconstruction: the observationally constrained background $H(z)$ determines $G_{\mathrm{eff}}(z)$
and, in turn, the geometric scalar profile that enters the subsequent matter-sector dynamics.

\section{Matter Scalar Sector : Higgs-like Potential and Variation of Vacuum Expectation Value}
One of the crucial components of our matter Lagrangian is a Higgs-like scalar field. The self-interaction of this field can provide the minimal realization of spontaneous symmetry breaking through the form \citep{Chakrabarti_2022, Chakrabarti2024},
\begin{equation}
V\left\lbrace\psi(z)\right\rbrace = V_0 + M(z)\psi^2 + \frac{\lambda}{4}\psi^4,
\end{equation}
where $\lambda$ is a constant self-coupling and $M(z)$ denotes an effective mass-term whose value is allowed to evolve with redshift. $\psi$ is regarded here as an effective cosmological Higgs-like scalar introduced to capture any cosmological symmetry-breaking or transition of phase. The analogy with the Standard Model Higgs sector is structural: in both cases, the quartic self-interaction together with the sign of the quadratic coefficient determines whether the potential can have a symmetry-preserving or symmetry-breaking configuration. The coefficient $M(z)$ is not fixed a priori and is subsequently determined from a cosmological reconstruction. For a given reconstructed potential $V(\psi)$, the corresponding effective mass coefficient can be extracted algebraically as
\begin{equation}
M(z) = \frac{1}{\psi^2}\left[V(\psi) - V_0 - \frac{\lambda}{4}\psi^4\right].
\end{equation}

We also express the scalar field and its effective potential in terms of the cosmological redshift. We further parametrize the mass coefficient as $M(z) \simeq M_0 u(z)$, where $M_0$ sets the characteristic mass scale of the matter-sector scalar and $u(z)$ describes its redshift dependence. The stationary configurations of the potential are obtained from
\begin{equation}
\frac{dV}{d\psi} = 2M(z)\psi + \lambda \psi^3 = 0.
\end{equation}
For $\lambda > 0$, this condition admits the trivial solution $\psi = 0$ and when $M(z) < 0$, two nontrivial minima. The latter correspond to the symmetry-breaking phase, for which the vacuum expectation value (VEV) of the scalar field is
\begin{equation}\label{delmu}
\nu(z) \equiv \langle \psi \rangle = \sqrt{-\frac{2M(z)}{\lambda}}, \qquad M(z) < 0.
\end{equation}
As a result, any cosmological evolution of the effective mass coefficient $M(z)$ induces a corresponding evolution of the symmetry-breaking scale $\nu(z)$. This provides a connection between the reconstructed scalar field and a cosmological evolution of particle masses. Assuming a constant Yukawa coupling $\lambda_{e,q}$, the fermion masses obey $m_{e,q} = \lambda_{e,q}\,\nu(z)$ \citep{CALMET2002173, Calmet_2017, GASSER198277, PhysRevLett.74.1071, PhysRevLett.121.212001}. The resulting evolution can be constrained by estimating the proton-to-electron mass ratio, $\mu \equiv m_p/m_e$, which is tightly constrained by spectroscopic observations of quasar absorption spectra \citep{PhysRevLett.113.123002}. The fractional variation of $\mu$ is defined as
\begin{equation}
\frac{\Delta \mu}{\mu} = \frac{\Delta m_p}{m_p} - \frac{\Delta m_e}{m_e}.
\end{equation}

The electron mass depends directly on the matter-sector VEV through the Yukawa interaction, whereas the proton mass is predominantly determined by the QCD confinement scale. Consequently, within the approximation adopted here, the dominant contribution to the variation of $\mu$ arises from the evolution of the electron mass. This hierarchy motivates the Lattice-QCD-based approximation \citep{GASSER198277, Ubachs_2017, Dent_2009}
\begin{equation}\label{delmumu}
\frac{\Delta \mu}{\mu} \simeq -\frac{91}{100}\,\frac{\Delta \nu}{\nu}.
\end{equation}
Once the effective mass function $M(z)$ is determined from the cosmological background within a generalized BD framework, the corresponding VEV
evolution $\nu(z)$ follows from Eq.~\eqref{delmu}, and the resulting $\Delta\mu/\mu$ can subsequently be derived and compared with observational bounds. In Table \ref{tab:deltamu}, we summarize the observational constraints on the variation of the proton-to-electron mass ratio. The table indicates that the variation is tightly bound, satisfying $|\Delta\mu/\mu| \approx 10^{-5} - 10^{-6}$ in Hubble scale \citep{PhysRevLett.113.123002}. 

\begin{table*}
\centering
\caption{Observational bounds on $\Delta\mu/\mu$, derived from high-resolution molecular absorption spectra }
\label{tab:deltamu}

\begin{tabular}{lcccc}
\hline
\textbf{Object} & \textbf{$z$} & $\boldsymbol{\Delta\mu/\mu~[10^{-6}]}$ & \textbf{Method} & \textbf{Ref.} \\
\hline

B0218+357  & 0.685 & $0.74 \pm 0.89$  & $NH_3/HCO^+/HCN$ & \citep{Murphy:2008yy} \\
B0218+357  & 0.685 & $-0.35 \pm 0.12$ & $NH_3/CS/H_2CO$  & \citep{Kanekar:2011yb} \\

PKS1830--211 & 0.886 & $0.08 \pm 0.47$  & $NH_3/HC_3N$ & \citep{Henkel_2009} \\
PKS1830--211 & 0.886 & $-1.2 \pm 4.5$   & $CH_3NH_2$   & \citep{ilyushin2012sensitivity} \\
PKS1830--211 & 0.886 & $-2.04 \pm 0.74$ & $NH_3$       & \citep{Muller_2011} \\
PKS1830--211 & 0.886 & $-0.10 \pm 0.13$ & $CH_3OH$     & \citep{Bagdonaite:2013sia} \\

J2123--005 & 2.059 & $8.5 \pm 4.2$ & $H_2/HD~(\mathrm{VLT})$  & \citep{van_Weerdenburg_2011} \\
J2123--005 & 2.059 & $5.6 \pm 6.2$ & $H_2/HD~(\mathrm{Keck})$ & \citep{Malec_2010} \\

HE0027--1836 & 2.402 & $-7.6 \pm 10.2$ & $H_2$ & \citep{Rahmani_2013} \\
Q2348--011  & 2.426 & $-6.8 \pm 27.8$ & $H_2$ & \citep{Bagdonaite_2012} \\
Q0405--443  & 2.597 & $10.1 \pm 6.2$  & $H_2$ & \citep{King_2008} \\
J0643--504  & 2.659 & $7.4 \pm 6.7$   & $H_2$ & \citep{AlbornozVasquez:2013wfw} \\
J1237+0648  & 2.688 & $-5.4 \pm 7.5$  & $H_2/HD$ & \citep{Dapr__2015} \\
Q0528--250  & 2.811 & $0.3 \pm 3.7$   & $H_2/HD$ & \citep{King_2011} \\
Q0347--383  & 3.025 & $2.1 \pm 6.0$   & $H_2$ & \citep{Wendt_2008} \\
J1443+2724  & 4.224 & $-9.5 \pm 7.6$  & $H_2$ & \citep{Bagdonaite_2015} \\
B0642--5038 & 2.66  & $10.3 \pm 4.6$  & $H_2$ & \citep{Bagdonaite_2014} \\

\hline
\end{tabular}
\end{table*}

\section{A Reconstruction based on Late-time Cosmic Expansion : Pade\'-Deformed $\Lambda$CDM}

\begin{figure*}
\centering
\includegraphics[width=\linewidth]{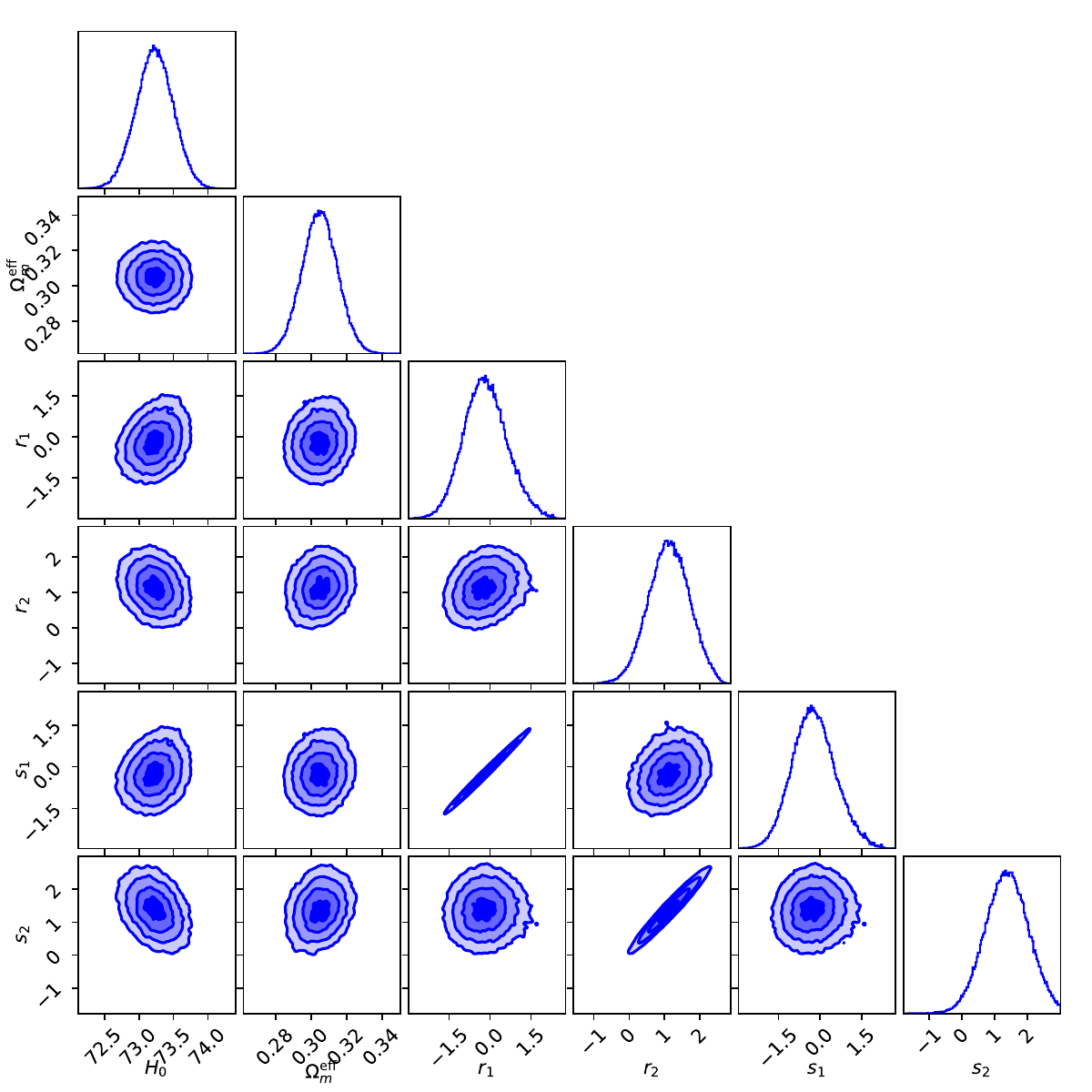}
\caption{The figure shows contours corresponding to $68\%$ and $95\%$ confidence levels, found from a Bayesian analysis using the combined data-set CC+BAO+DESI+Pantheon+SH0ES+CMB.}
\label{Corner2}
\end{figure*}

\begin{figure}
\begin{center}
\includegraphics[width=0.40\textwidth]{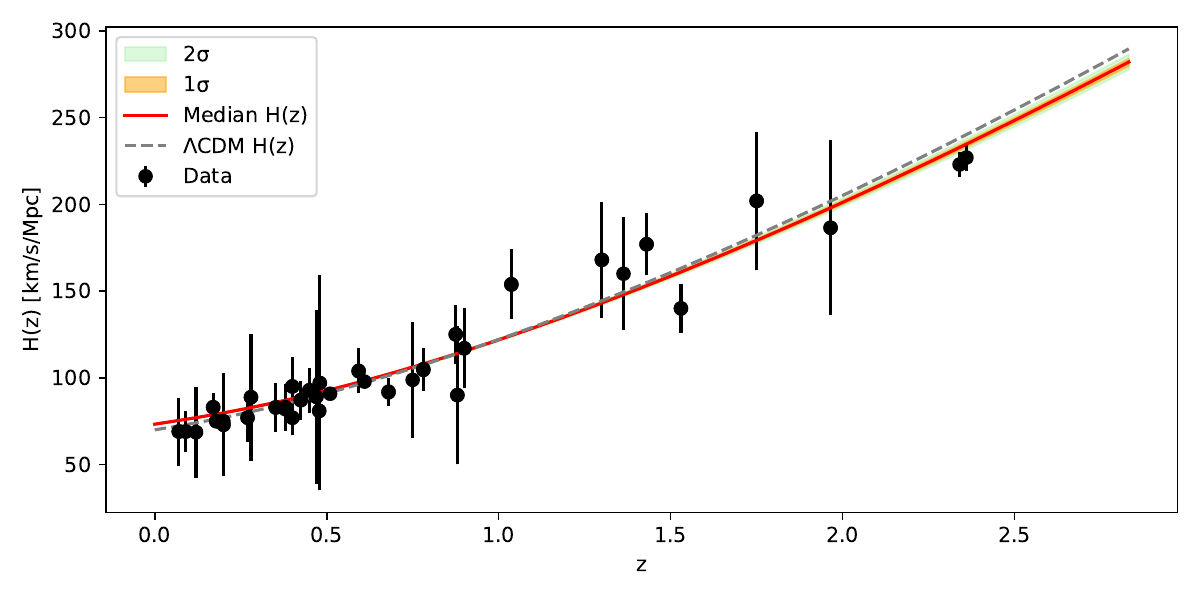}
\includegraphics[width=0.40\textwidth]{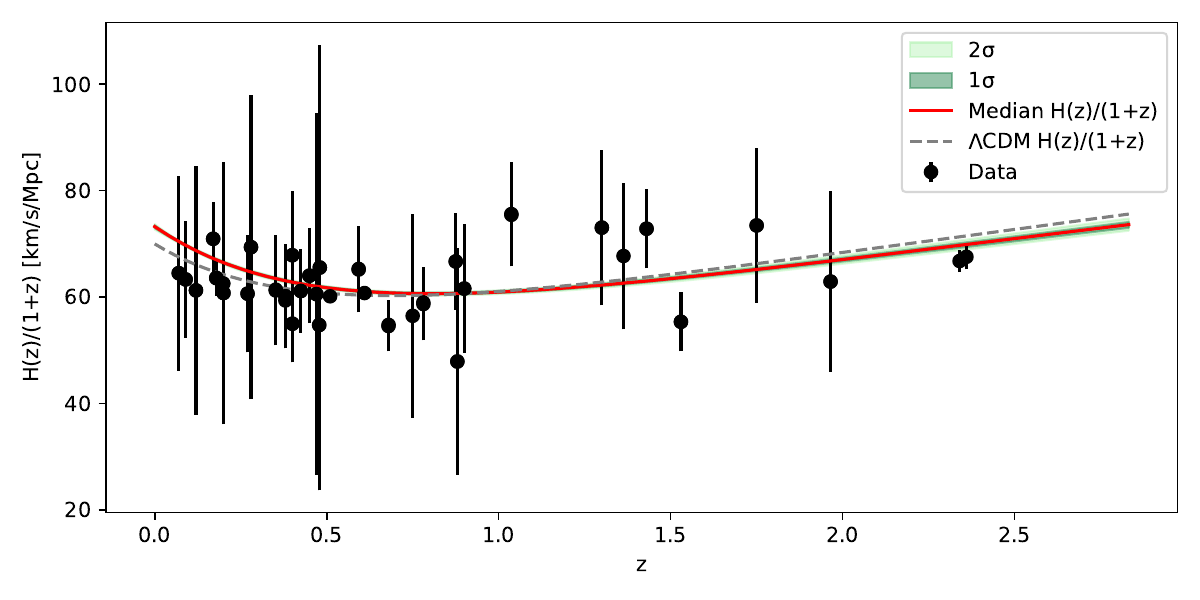}
\includegraphics[width=0.40\textwidth]{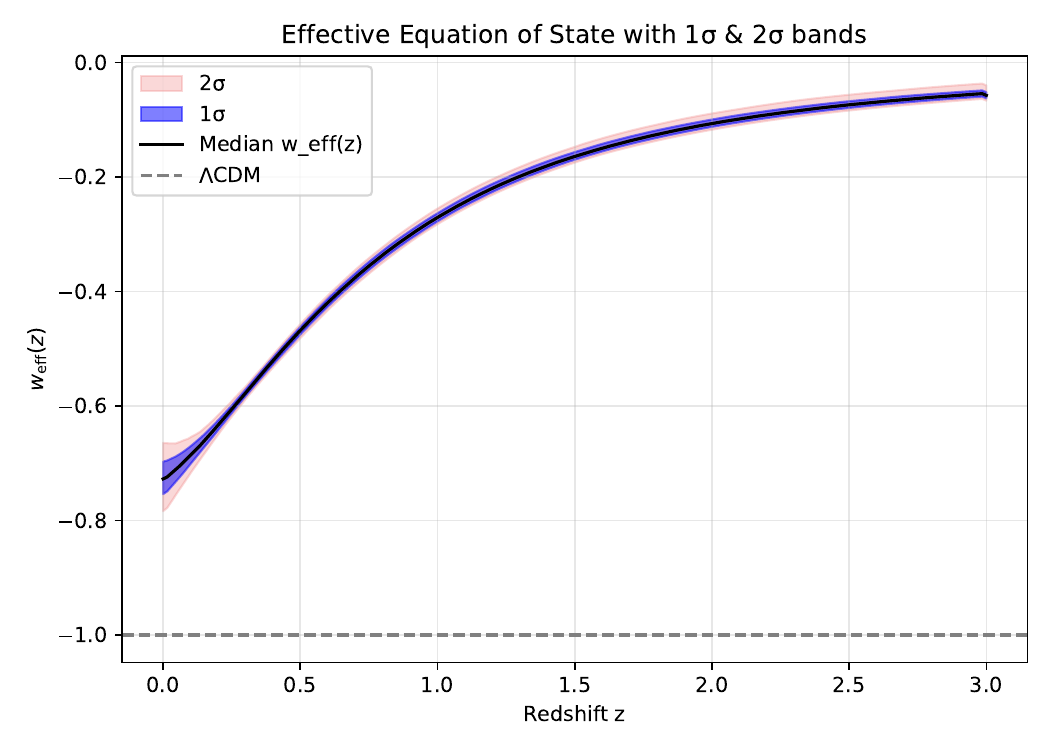}
\caption{Expansion history of the modified $\Lambda$CDM model from the combined CC+BAO+DESI+CMB data set. The top, middle, and bottom panels show $H(z)$, $H(z)/(1+z)$, and the effective equation-of-state parameter $\omega_{\rm eff}(z)$, respectively. The shaded areas represent the $1\sigma$ and $2\sigma$ confidence intervals.}
\label{Hubble1}
\end{center}
\end{figure}

We use a late-time cosmological expansion history as the background input for the reconstruction. The $\Lambda$CDM model provides a successful phenomenological description of a wide range of cosmological observations, including cosmic microwave background anisotropies, baryon acoustic oscillations, Type Ia supernovae, and large-scale structure formation \citep{Upadhye_2005}. In this framework, the late-time energy budget is dominated by cold dark matter and a cosmological constant $\Lambda$, with the latter responsible for the observed accelerated expansion. Despite its observational success, $\Lambda$CDM leaves open several conceptual questions, including the cosmological constant problem
\citep{RevModPhys.61.1} and the coincidence problem \citep{Velten_2014, PhysRevD.78.021302}. These considerations motivate the investigation of controlled departures from the standard expansion history while retaining $\Lambda$CDM as a well-defined limiting case. Several phenomenological parametrizations have been proposed to describe departures from $\Lambda$CDM, including the Chevallier-Polarski-Linder (CPL) parametrization \citep{Chevallier:2000qy, PhysRevLett.90.091301, Linden:2008mf}, alternative two-parameter forms \citep{Jassal:2004ej, PhysRevD.72.103503, Barboza:2008rh, PhysRevD.77.123525, Wetterich:2004pv}, and Quintom scenarios \citep{Feng:2004ad, Guo:2004fq}. In the present work, we adopt a rational Pad\'e-deformation of the $\Lambda$CDM Hubble function,
\begin{equation}\label{Hpade}
H(z) = H_0\sqrt{\Omega_m^{\rm eff}(1+z)^3 + 1-\Omega_m^{\rm eff}} \left\{\frac{1+r_1u+r_2u^2}{1+s_1u+s_2u^2} \right\},
\end{equation}
where $u=\frac{1}{1+z}-1$. The coefficients $(r_1,r_2,s_1,s_2)$ parameterize deviations from the reference $\Lambda$CDM expansion, which is recovered when $r_1 = r_2 = s_1 = s_2 = 0$. The Pad\'e form provides a convenient phenomenological extension in which departures from the standard background are introduced in a controlled manner without abandoning the $\Lambda$CDM limit. We constrain the parameters of Eq.~\eqref{Hpade} using a Bayesian analysis that combines complementary cosmological probes, including cosmic chronometers (CC), baryon acoustic oscillations (BAO), Type Ia supernovae (SNe Ia), and cosmic microwave background (CMB) distance information. The posterior distribution is sampled using the Markov Chain Monte Carlo (MCMC) ensemble sampler implemented in the \texttt{emcee} package \citep{ForemanMackey2013}. The parameter space is six-dimensional,
\[
\boldsymbol{\theta}
=
\left(
H_0,\Omega_m^{\rm eff},r_1,r_2,s_1,s_2
\right),
\]
and standard convergence diagnostics are employed to verify the stability of the posterior sampling.  \\

We use the following data-sets : (i) Cosmic chronometer measurements which provide direct estimates of the Hubble parameter through the differential age evolution of passively evolving galaxies \citep{Jimenez2003, Moresco2016, Moresco:2020fbm}. Their relatively weak dependence on an assumed background cosmology makes them particularly useful for constraining the expansion rate directly. (ii) We additionally include baryon acoustic oscillation measurements, which provide distance information through the relic imprint of primordial acoustic oscillations. In our analysis, these include measurements from SDSS, BOSS, and the DESI DR2 release \citep{Eisenstein_2005, Anderson2014, DESI2025}. (iii) We include Type-Ia supernovae data through the Pantheon + SH0ES compilation \citep{Brout2022}, which constrains the relative luminosity-distance evolution. Early-Universe geometric information is included through the Planck 2018 CMB distance priors \citep{Planck2018}. The resulting posterior constraints are summarized in Table \ref{tab:mcmc_results}. For the combined data set considered here, we obtain
\[
H_0 = 73.22 \pm 0.26~{\rm km\,s^{-1}\,Mpc^{-1}}.
\]
We do not interpret this result by itself as evidence for a resolution of the $H_0$ tension. In particular, the inferred value can be sensitive to the specific combination of data sets adopted here, especially the inclusion of the Pantheon+SH0ES compilation. We therefore regard the reconstructed value of $H_0$ primarily as the posterior constraint associated with the present data combination, rather than as evidence for a definitive alleviation of the $H_0$ tension. \\

\begin{table}
\centering
\caption{Posterior constraints on the parameters of the six-parameter
Pad\'e-deformed Hubble expansion obtained from the combined cosmological
data set. The quoted uncertainties correspond to the $1\sigma$ credible
intervals.}
\begin{tabular}{lc}
\hline
Parameter & Constraint \\
\hline
$H_{0}$ & $73.22 \pm 0.26$ \\
$\Omega_{m}^{\rm eff}$ &
$0.3049^{+0.0098}_{-0.0100}$ \\
$r_{1}$ &
$-0.184^{+0.803}_{-0.747}$ \\
$r_{2}$ &
$1.134^{+0.575}_{-0.581}$ \\
$s_{1}$ &
$-0.238^{+0.788}_{-0.732}$ \\
$s_{2}$ &
$0.698^{+0.649}_{-0.642}$ \\
\hline
\end{tabular}
\label{tab:mcmc_results}
\end{table}

The statistical preference of the Pad\'e-deformed expansion over the reference $\Lambda$CDM model is assessed using the differences in the
chi-square, Akaike information criterion (AIC), corrected Akaike information criterion (AICc), and Bayesian information criterion (BIC).
For two competing models, we define
\begin{equation}
\Delta X \equiv X_{\rm Pad\acute{e}}-X_{\Lambda{\rm CDM}},
\end{equation}
so that negative values favor the Pad\'e model. The chi-square difference is defined as
\begin{equation}
\Delta\chi^2 = \chi^2_{\rm Pad\acute{e}} - \chi^2_{\Lambda{\rm CDM}}.
\end{equation}
For a model with $k$ free parameters, the Akaike information criterion is defined as
\begin{equation}
\mathrm{AIC} = \chi^2_{\min}+2k,
\end{equation}
where $\chi^2_{\min}$ denotes the minimum chi-square. Since the AIC can be biased when the number of fitted parameters is not negligible compared with the number of data points, we also consider the finite-sample corrected form,
\begin{equation}
\mathrm{AICc} = \mathrm{AIC} + \frac{2k(k+1)}{N-k-1},
\end{equation}
where $N$ is the effective number of data points used in the fit. The Bayesian information criterion is given by
\begin{equation}
\mathrm{BIC} = \chi^2_{\min}+k\ln N.
\end{equation}
Both AIC and BIC penalize additional model parameters, with BIC imposing a stronger penalty for sufficiently large data sets. The corresponding differences are therefore
\begin{align}
\Delta\mathrm{AIC}
&=
\mathrm{AIC}_{\rm Pad\acute{e}}
-
\mathrm{AIC}_{\Lambda{\rm CDM}},\\
\Delta\mathrm{AICc}
&=
\mathrm{AICc}_{\rm Pad\acute{e}}
-
\mathrm{AICc}_{\Lambda{\rm CDM}},\\
\Delta\mathrm{BIC}
&=
\mathrm{BIC}_{\rm Pad\acute{e}}
-
\mathrm{BIC}_{\Lambda{\rm CDM}}.
\end{align}

For the present data combination, we find that $\Delta\chi^2=-7.453$, showing that the additional freedom in the reconstructed expansion history
provides a noticeably better maximum-likelihood fit than $\Lambda$CDM. However, once the additional model parameters are penalized, this improvement is largely compensated. In particular, $\Delta{\rm AIC}=0.547$, which is sufficiently small to indicate that the two models have essentially comparable support according to the Akaike criterion. Therefore, although the extended model improves the raw goodness of fit, the data-set does not provide a statistically significant AIC preference for it over $\Lambda$CDM. The Bayesian information criterion gives a substantially larger value $\Delta{\rm BIC}=22.503$, and consequently favors the more economical $\Lambda$CDM description. The difference between the AIC and BIC conclusions is expected because the BIC applies a considerably stronger complexity penalty for the large number of data points, even more so, after the inclusion of the Pantheon+SH0ES sample. \\

The reconstructed expansion history is shown in Fig.~\ref{Hubble1} which suggests that $H(z)$ remains close to the $\Lambda$CDM behavior, while permitting modest low-redshift departures. The corresponding effective equation-of-state parameter $\omega_{\rm eff}(z)$ displays a smooth evolution and a natural transition from matter-dominated epoch into a late-time accelerated regime.

\section{Reconstruction of the Generalized Brans-Dicke Lagrangian}
Once the cosmological expansion history is specified, the scalar-field reconstruction can be executed. We consider two frameworks under the broad class of generalized Brans-Dicke theory. In both of the cases, the matter-sector scalar $\psi$ is minimally coupled and is responsible for the evolution of the symmetry-breaking scale. The two models differ only in the coupling of the geometric Brans-Dicke scalar $\phi$ to the matter sector: in the first case this coupling is absent, whereas in the second case the geometric scalar interacts with Baryonic matter, i.e., the interaction is promoted to a chameleon-type environment-dependent interaction. 

\subsection{Running-vacuum scalar with a minimally coupled Brans-Dicke field}
\label{sec:BD_RV}
For the framework where a matter-sector scalar $\psi$ is minimally coupled to the standard Brans-Dicke action, the Jordan-frame action is
given by
\begin{equation}
S = \int d^4x \sqrt{-g} \left[\phi R + \frac{\omega}{\phi}\partial_\mu\phi\,\partial^\mu\phi + L_\psi \right],
\end{equation}
where $R$ is the Ricci scalar, $g$ is the determinant of the metric tensor $g_{\mu\nu}$, $\phi(t)$ is the Brans-Dicke scalar, $\omega$ is the
dimensionless Brans-Dicke parameter and $L_\psi$ is the matter-sector Lagrangian. The corresponding cosmological field equations are derived as
\begin{eqnarray}\label{fe1BD}
&& 3 \frac{\dot{a}^2}{a^2} = \frac{\rho_\psi}{\phi} + \frac{\omega}{2}\frac{\dot{\phi}^2}{\phi^2} - 3\frac{\dot{a}}{a}\frac{\dot{\phi}}{\phi},\\&&\label{fe2BD}
2 \frac{\ddot{a}}{a} + \frac{\dot{a}^2}{a^2} = -\frac{\omega}{2}\frac{\dot{\phi}^2}{\phi^2} - \frac{\ddot{\phi}}{\phi} - 2\frac{\dot{a}}{a}\frac{\dot{\phi}}{\phi} - \frac{p_\psi}{\phi},\\&&
\rho_\psi = \frac{1}{2}\dot{\psi}^{\,2}+V(\psi),~~p_\psi = \frac{1}{2}\dot{\psi}^{\,2}-V(\psi).
\end{eqnarray}
The equation of motion for $\psi$ takes the standard Klein-Gordon form
\begin{equation}\label{fe3BD}
\ddot{\psi} + 3\frac{\dot{a}}{a}\dot{\psi} + \frac{dV}{d\psi} = 0.
\end{equation}
The Brans-Dicke scalar equation following from the action is
\begin{equation}\label{fe4BD}
\ddot{\phi} + 3\frac{\dot{a}}{a}\dot{\phi} = \frac{4V(\psi)-\dot{\psi}^{\,2}}{2\omega+3}.
\end{equation}

The matter-sector potential is taken in the effective Higgs-like form as discussed before, i.e., $V(\psi) = V_0 + M(z)\psi^2 + \frac{\lambda}{4}\psi^4$. The redshift-dependent coefficient $M(z)$ is not assumed at the outset and will be derived as part of the cosmological reconstruction. 
The connection between the BD scalar and the effective gravitational coupling is imposed through the RG-motivated running as in Eq. \eqref{eq:runningG}. For the reconstruction it is convenient to express the field equations entirely in terms of redshift. Using $1+z=\frac{1}{a}$, the cosmic-time derivative can be replaced by
\begin{equation}
\frac{d}{dt} = -(1+z)H(z)\frac{d}{dz}.
\end{equation}
Consequently, for any scalar quantity $X(z)$, $\dot{X} = -(1+z)H X_{,z}$ and $\ddot{X} = (1+z)^2H^2X_{,zz} + (1+z)H \left[H+(1+z)H_{,z}\right]X_{,z}$. We write Eqs.~\eqref{fe1BD}-\eqref{fe4BD} in terms of $z$, $H(z)$, and the redshift derivatives of the scalar fields.   \\

\begin{figure}
\begin{center}
    \includegraphics[width=0.4\textwidth]{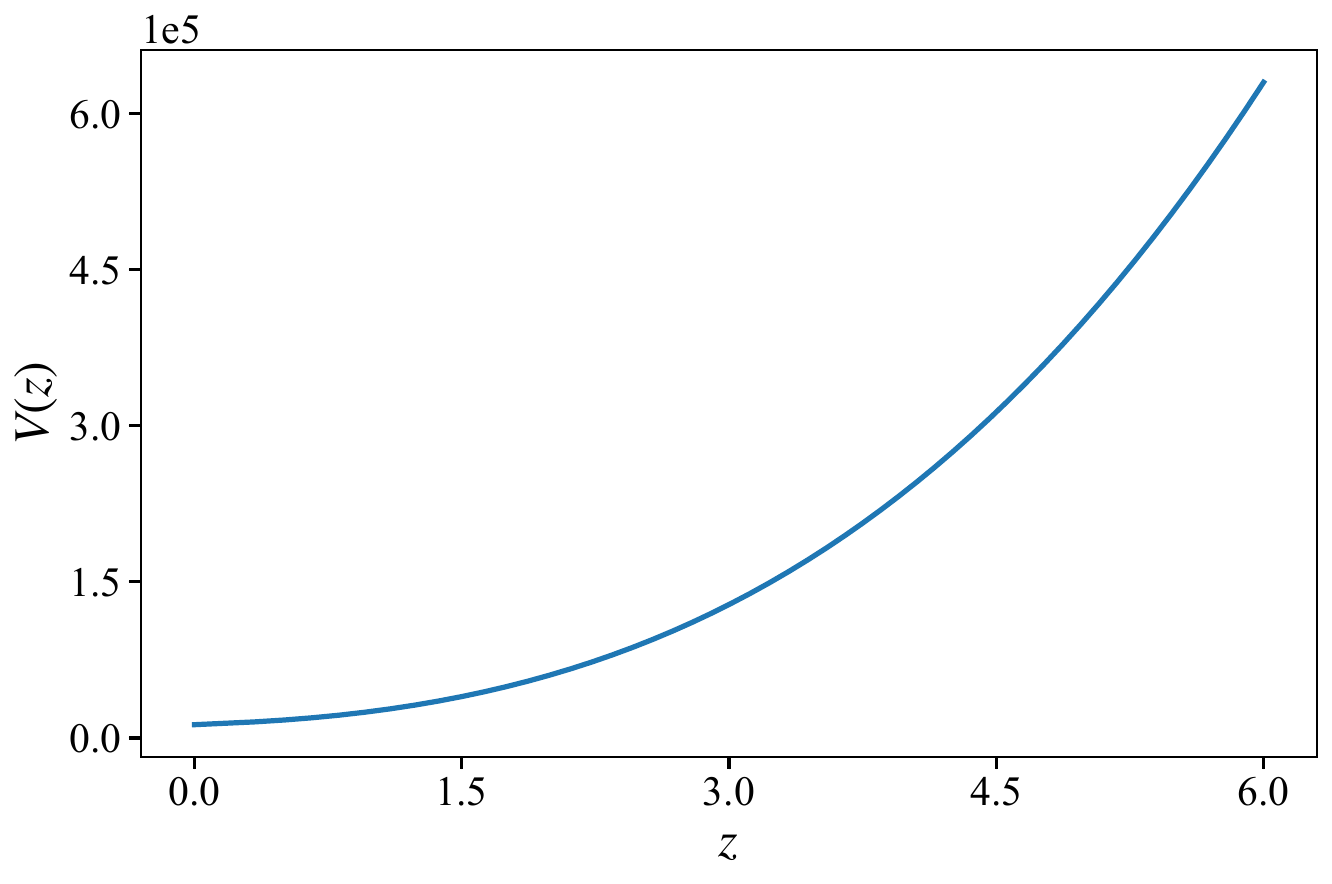}
    \includegraphics[width=0.4\textwidth]{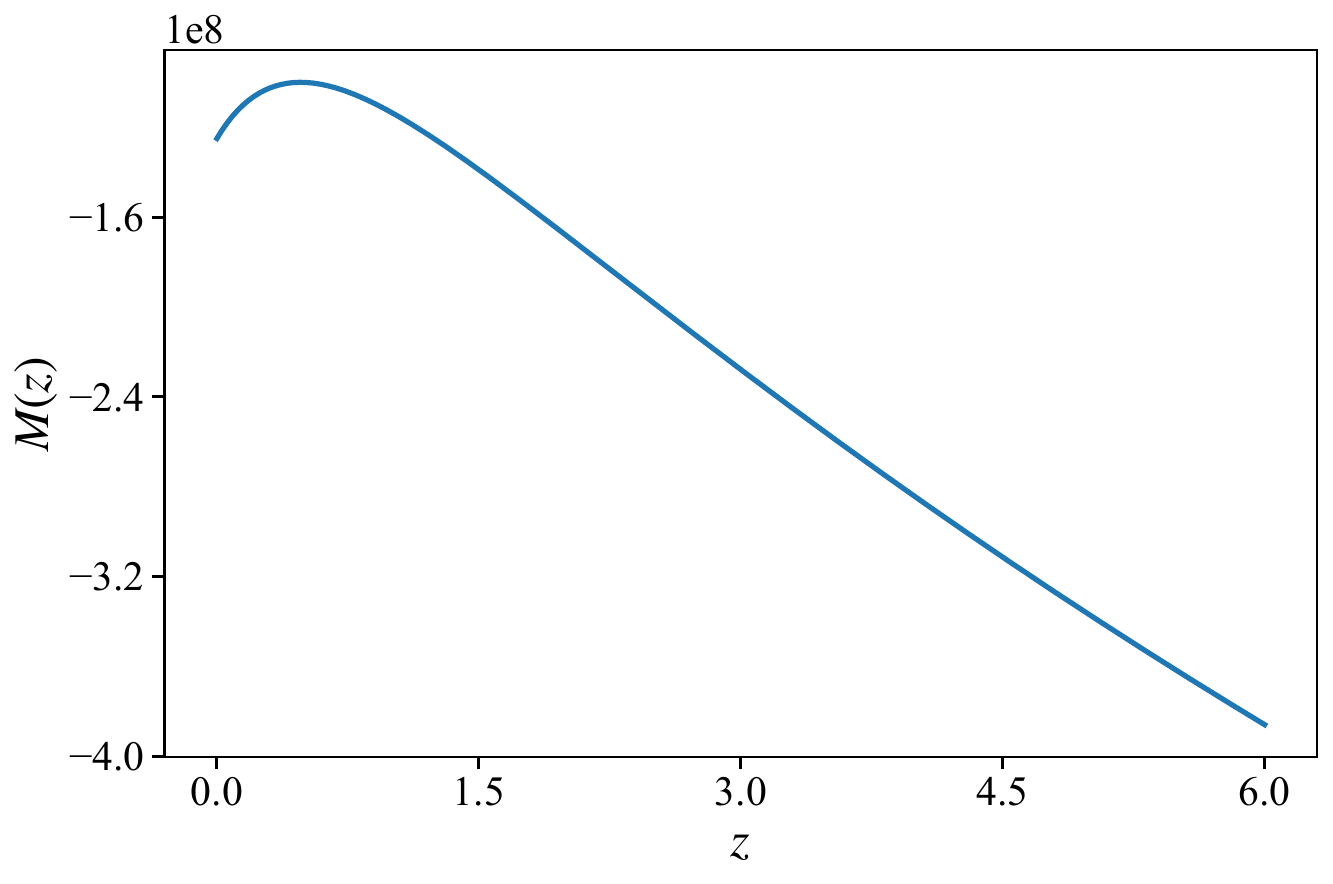}
    \includegraphics[width=0.4\textwidth]{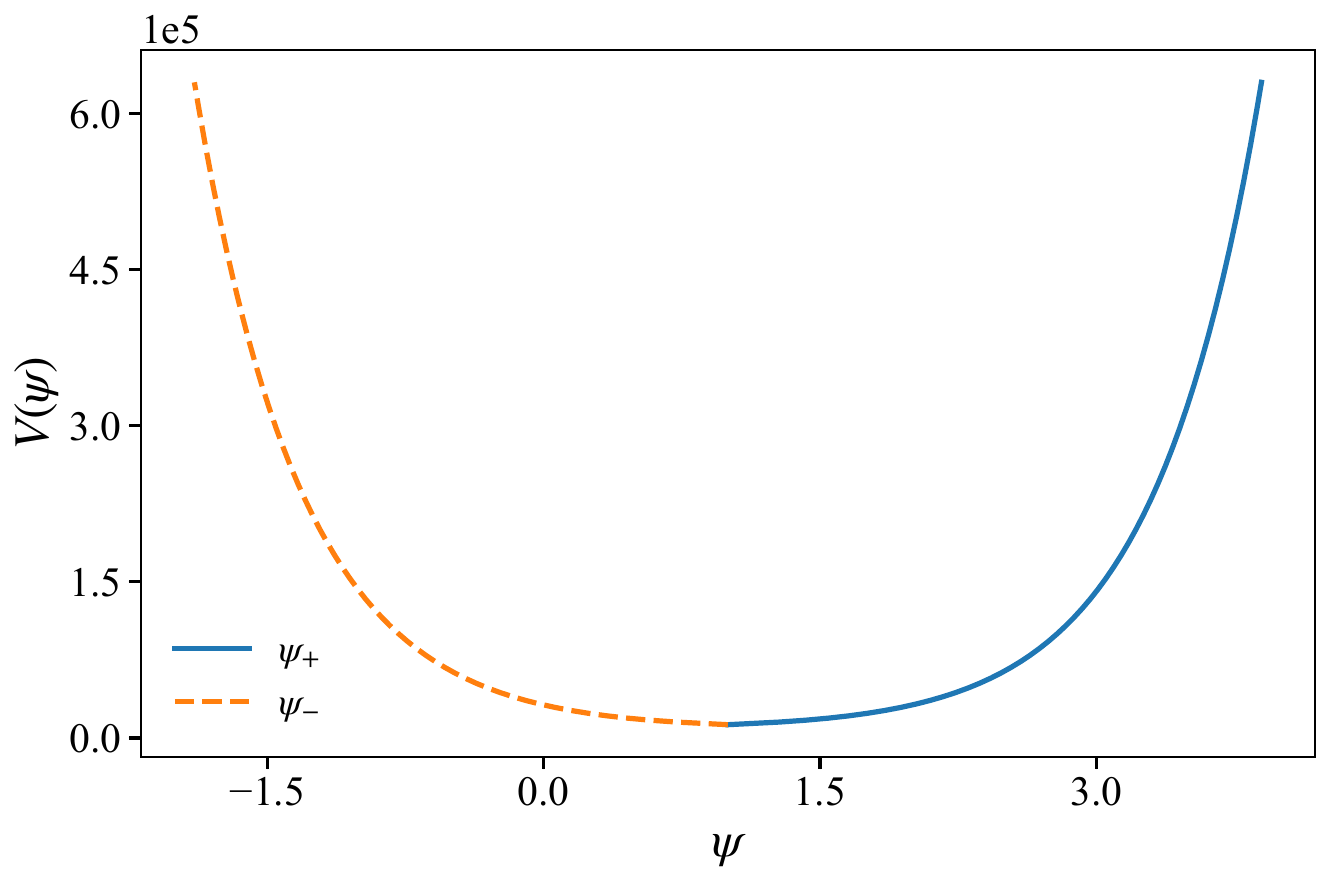}
    \caption{Reconstructed scalar potential $V(z)$, evolution of effective mass $M(z)$ with redshift and parametric reconstruction $V(\psi)$ for both branches of $\psi$ in generalized Brans--Dicke framework respectively.}
    \label{fig11}
\end{center} 
\end{figure}

\begin{figure}
\begin{center}
    \includegraphics[width=0.4\textwidth]{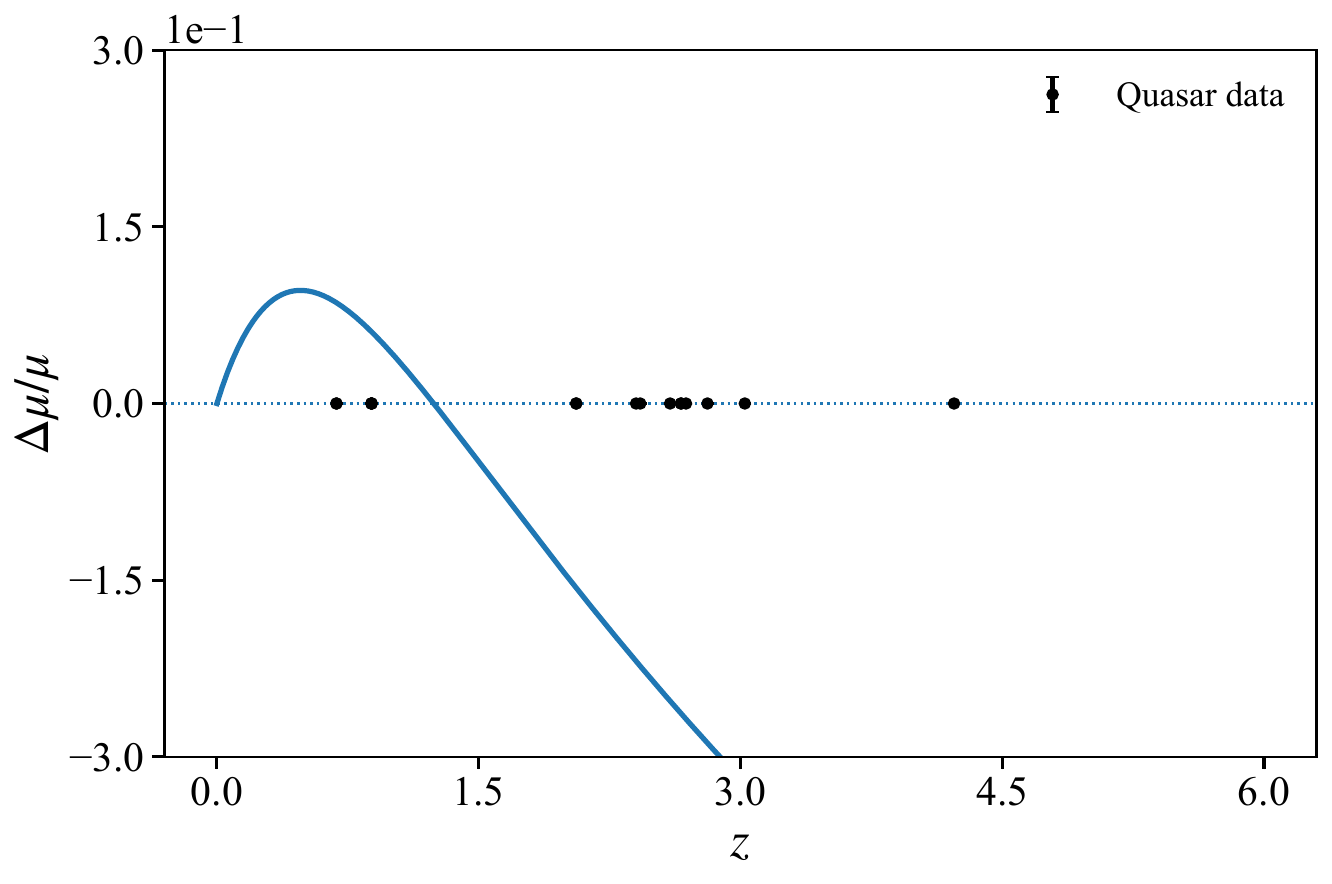}    
   \caption{Variation of the proton-to-electron mass ratio $\Delta\mu/\mu$, with redshift in generalized Brans--Dicke framework.}
    \label{fig111}
\end{center}
\end{figure}

Throughout the reconstruction, the effective mass coefficient is normalized according to $M(z)=M_0 u(z)$, where $M_0\sim\mathcal{O}(H_0^2)$ sets the characteristic mass scale and $u(z)$ contains the redshift dependence determined by the scalar-field dynamics. The background Hubble expansion is specified by the Pad\'e-inspired form in Eq.~\eqref{Hpade}. The $\Lambda$CDM expansion is contained as the limiting case $r_1 = r_2 = s_1 = s_2 = 0$ and therefore does not need to be treated as a separate reconstruction.  \\

The system of equations is solved numerically. The reconstruction gives us the redshift evolution of the matter-sector scalar $\psi(z)$, the effective potential $V(z)$, and the mass coefficient $M(z)$. The potential can subsequently be reconstructed parametrically as $V(\psi)$. The resulting profiles are shown in Fig.~\ref{fig11}. The evolution of $M(z)$ determines the corresponding vacuum expectation value through Eq.~\eqref{delmu}, which gives the predicted variation of the proton-to-electron mass ratio through Eq.~\eqref{delmumu}. The resulting $\Delta\mu/\mu$ is shown in Fig.~\ref{fig111} together with the observational data-points listed in Table~\ref{tab:deltamu}. The figure suggests that the variation of $\Delta\mu/\mu$ is too large compared to the observationally allowed variation.

\subsection{Chameleon Brans--Dicke Extension}
\label{sec:chameleonBD}

We now extend the generalized Brans-Dicke framework by allowing the geometric scalar $\phi$ to couple non-minimally to ordinary matter through
a function $f(\phi)$. The matter-sector scalar $\psi$ is retained with minimal coupling, as in the preceding reconstruction, and continues to
control the evolution of the vacuum expectation value. Thus, the two models differ only in the coupling of the geometric scalar to ordinary matter. This coupling is characteristic of chameleon-like theories and makes the scalar interaction sensitive to the local density \cite{PhysRevD.75.104016, Harko2010, Faraoni2004, PhysRevD.78.043512, Chakrabarti_2022}. We intend to examine whether this additional coupling can modify the reconstructed evolution of the matter-sector field enough to control the cosmological variation of the proton-to-electron mass ratio.
\\

The Jordan-frame action is taken to be
\begin{equation}
S = \int d^4 x \sqrt{-g}\left[ \phi R - \frac{\omega}{\phi} g^{\mu\nu}\partial_\mu\phi\,\partial_\nu\phi +2L_\psi +2f(\phi)L_m \right].
\label{action}
\end{equation}
Here, $L_m$ denotes the ordinary matter Lagrangian, while $L_\psi$ describes the minimally coupled matter-sector scalar as introduced before. The choice of matter Lagrangian is relevant once a non-minimal scalar-matter coupling is present. In minimally coupled theories, different fluid representations such as $L_m=-\rho_m$ and $L_m=p_m$ are dynamically equivalent at the level of the gravitational equations. However, they need not remain equivalent when multiplied by a field-dependent function $f(\phi)$ \cite{PhysRevD.75.104016}. Following the prescription adopted in Das and Banerjee \cite{PhysRevD.78.043512}, we take
\begin{equation}
L_m=-\rho_m.
\end{equation}
With this choice, variation of the action with respect to the metric gives
\begin{align}
\phi G_{\mu\nu} &=
f(\phi)T^{(m)}_{\mu\nu}
+T^{(\psi)}_{\mu\nu}
+\frac{\omega}{\phi}
\left(
\nabla_\mu\phi\nabla_\nu\phi
-\frac{1}{2}g_{\mu\nu}(\nabla\phi)^2
\right)
\nonumber\\
&\quad
+\nabla_\mu\nabla_\nu\phi
-g_{\mu\nu}\Box\phi .
\label{einstein-eq}
\end{align}
The equation governing the geometric scalar is
\begin{equation}
(2\omega+3)\Box\phi = f(\phi)T^{(m)} + T^{(\psi)} - 2\phi f'(\phi)L_m,
\label{phi-eq}
\end{equation}
where $f'(\phi)=df/d\phi$. The non-minimal coupling also modifies the matter-sector conservation equation and gives
\begin{equation}
\nabla^\mu T^{(m)}_{\mu\nu} = \frac{f'(\phi)}{f(\phi)} \left(L_m g_{\mu\nu}-T^{(m)}_{\mu\nu} \right) \nabla^\mu\phi.
\label{noncons}
\end{equation}
This means that the ordinary matter stress tensor is not separately conserved in the presence of a field-dependent coupling. The precise form of the resulting exchange depends on the matter Lagrangian adopted. For a homogeneous and isotropic background, defining $\rho_\psi=\frac{1}{2}\dot{\psi}^{\,2}+V(\psi)$ and $p_\psi=\frac{1}{2}\dot{\psi}^{\,2}-V(\psi)$, we derive the modified Friedmann equations as
\begin{align}
3H^2
&=
\frac{f(\phi)}{\phi}\rho_m
+\frac{\rho_\psi}{\phi}
+\frac{\omega}{2}
\left(\frac{\dot{\phi}}{\phi}\right)^2
-3H\frac{\dot{\phi}}{\phi},
\label{fried1}
\\[4pt]
2\dot H+3H^2
&=
-\frac{f(\phi)}{\phi}p_m
-\frac{p_\psi}{\phi}
-\frac{\omega}{2}
\left(\frac{\dot{\phi}}{\phi}\right)^2
-2H\frac{\dot{\phi}}{\phi}
-\frac{\ddot{\phi}}{\phi}.
\label{fried2}
\end{align}
The equation of motion of $\psi$ retains the standard Klein-Gordon form,
\begin{equation}
\ddot{\psi}+3H\dot{\psi}+V'(\psi)=0.
\label{psi-eq}
\end{equation}
The corresponding equation for the geometric scalar is
\begin{equation}
(2\omega + 3) \left(\ddot{\phi}+3H\dot{\phi}\right) = f(\rho_m-3p_m) - \dot{\psi}^{\,2} + 4 V -2\phi f'\rho_m .
\label{phi-FRW}
\end{equation}

It is useful to note here, that the function $f(\phi)$ does not enter Eq. \eqref{psi-eq} directly; instead, its effect on $\psi$ is mediated through the modified evolution of the geometric scalar $\phi$ and the resulting cosmological background. The terms proportional to $f$ and $f'$ in Eq. \eqref{phi-FRW} alter the source driving $\phi$, which subsequently changes the reconstruction of the matter-sector quantities $\psi(z)$, $M(z)$, and $V(z)$. Thus, any suppression of the predicted variation of $\mu$ is a consequence of the coupled cosmological solution,
rather than a direct modification of the $\psi$ equation. Combining Eqs. \eqref{fried1}-\eqref{phi-FRW}, the kinetic term and potential of the matter-sector scalar can be reconstructed as
\begin{align}
\dot{\psi}^{\,2}
&=
\frac{1}{3}\Bigg[
4\left(
3H^2\phi
-f\rho_m
-\frac{\omega}{2}\frac{\dot{\phi}^{\,2}}{\phi}
+3H\dot{\phi}
\right)
\nonumber\\
&\quad
-(2\omega+3)
\left(
\ddot{\phi}+3H\dot{\phi}
\right)
+f(\rho_m-3p_m)
-2\phi f'\rho_m
\Bigg],
\\[1em]
V(\psi)
&=
\frac{1}{6}\Bigg[
2\left(
3H^2\phi
-f\rho_m
-\frac{\omega}{2}\frac{\dot{\phi}^{\,2}}{\phi}
+3H\dot{\phi}
\right)
\nonumber\\
&\quad
+(2\omega+3)
\left(
\ddot{\phi}+3H\dot{\phi}
\right)
-f(\rho_m-3p_m)
+2\phi f'\rho_m
\Bigg].
\label{Vpsi}
\end{align}

\subsubsection{Chameleon coupling, screening, and local-gravity constraints}
\label{sec:chameleon_constraints}
In this stage it is crucial to choose the scalar-matter coupling, which is a central ingredient of the chameleon extension. In particular, a viable screening mechanism requires the scalar-mediated interaction to be sufficiently suppressed in high-density regions while allowing the
scalar to remain dynamically relevant on cosmological scales \cite{Khoury_2004, PhysRevD.75.104016, Harko2010}. The chameleon mechanism provides such a realization through the density-dependent effective potential of the scalar field. We briefly review the construction and the resulting constraints included in the screening analysis developed in \citep{Chakrabarti_2022}. Starting from the generalized Chameleon Brans-Dicke action in the Jordan frame,
\begin{equation}
S=\frac{1}{16\pi}\int d^4x\sqrt{-\bar g}\left[ \phi\bar R-\frac{\omega_{\rm BD}}{\phi} \bar g^{\mu\nu}\bar\nabla_\mu\phi\bar\nabla_\nu\phi
-V(\phi)+16\pi f(\phi)L_m \right],
\label{cham_action}
\end{equation}
we transform it to the Einstein frame choosing $g_{\mu\nu}=\Omega^2\bar g_{\mu\nu}$, $\Omega=\sqrt{G\phi}$ and introducing the canonically normalized scalar $\varphi(\phi) = \sqrt{\frac{2\omega_{\rm BD}+3}{16\pi G}} \ln\left(\frac{\phi}{\phi_0}\right)$, provided $\qquad \omega_{\rm BD}>-\frac{3}{2}$. The resulting Einstein-frame action is found as
\begin{equation}
S_{\rm EF} = \int d^4x\sqrt{-g} \left[\frac{R}{16\pi G} - \frac12(\nabla\varphi)^2 - U(\varphi) + h(\varphi)L_m \right],
\label{EF_action_cham}
\end{equation}
where $h(\varphi) = e^{-\sigma\varphi/M_p}f(\varphi)$, $\sigma = 8\sqrt{\frac{\pi}{2\omega_{\rm BD}+3}}$ and $U(\varphi) = V\bigl(\phi(\varphi)\bigr)e^{-\sigma\varphi/M_p}$. Thus, the Brans-Dicke parameter enters the matter coupling through $\sigma$, and constraints on the effective scalar-matter coupling translate directly into constraints on $\omega_{\rm BD}$. Variation of Eq. \eqref{EF_action_cham} with respect to $\varphi$ gives
\begin{equation}
\Box\varphi = U_{,\varphi} - h_{,\varphi}L_m.
\label{cham_eom}
\end{equation}
For a matter Lagrangian choice of $L_m=-\rho_m$, the matter contribution can be absorbed into an effective potential. In an environment of homogeneous density, this may be expressed as
\begin{equation}
U_{\rm eff}(\varphi) = U(\varphi) - \rho_0 \int \left(\frac{\sigma}{M_p} - \frac{df}{d\varphi}\right) e^{\sigma\varphi/(2M_p)}d\varphi ,
\label{effective_potential_cham}
\end{equation}
with the normalization of $\rho_0$ depending on the Einstein-frame matter content. The important feature is that the minima of $U_{\rm eff}$ and the corresponding fluctuation mass, depend on the ambient matter density,
\begin{equation}
U_{{\rm eff},\varphi}(\varphi_{\rm min})=0,~~ m_{\rm min}^2 = U_{{\rm eff},\varphi\varphi}(\varphi_{\rm min}).
\end{equation}
As a result, in high-density regions the scalar can acquire a large effective mass and remain close to its local minimum, whereas in a low-density environment its range can become much larger. This density dependence is the origin of the chameleon screening mechanism.  \medskip

For a spherical body of radius $R$ and density $\rho_c$ embedded in an environment of density $\rho_\infty$, the scalar equation takes the
form
\begin{eqnarray}\label{cham_spherical}
&& \frac{d^2\varphi}{dr^2} + \frac{2}{r}\frac{d\varphi}{dr} = U_{,\varphi} + \rho_0(r)\frac{\beta(\varphi)}{M_p}e^{\sigma\varphi/(2M_p)},\\&&
\frac{\beta(\varphi)}{M_p} = \frac{df}{d\varphi} - \frac{\sigma}{M_p}~,~\rho_{0} (r) = \left\{
\begin{matrix}
\rho_c~~ {\rm for}\;r < R \cr
\rho_\infty~~ {\rm for}\; r > R
\end{matrix}
\right. \,.\label{beta_general}
\end{eqnarray}
The quantity $\beta(\varphi)$ contains the explicit matter coupling $f(\varphi)$ as well as the coupling inherited from the Brans-Dicke transformation. For a sufficiently large body, the scalar remains close to the density-dependent minima throughout most of its interior and varies appreciably only within a thin layer near the surface. The resulting exterior profile can be derived as
\begin{equation}
\varphi(r) \simeq -\frac{\beta_0}{4\pi M_p}\left(\frac{3\Delta R}{R}\right) \frac{M_c e^{-m_\infty(r-R)}}{r} +\varphi_\infty ,
\label{thin_shell_profile}
\end{equation}
where $\beta_0$ denotes the approximately constant effective coupling and $\Delta R/R$ is the fractional thickness of the shell. The
thin-shell screening requires
\begin{equation}
\frac{\Delta R}{R} = \frac{M_p(\varphi_\infty-\varphi_c)}{\beta_0\rho_cR^2} \ll1.
\label{thin_shell_condition}
\end{equation}
For an unscreened object, the corresponding exterior solution lacks this suppression factor. The factor $\Delta R/R$ is directly responsible for suppressing the scalar generated force. The same mechanism can be applied to the Earth atmosphere system or from the differential free fall in the Earth-Moon-Sun system. Lunar laser-ranging measurements require
\begin{equation}
\frac{|a_{\rm Moon}-a_\oplus|}{a_N} \lesssim 10^{-13},
\label{EP_constraint}
\end{equation}
where $a_N$ is the Newtonian acceleration. For screened Earth and Moon profiles, the scalar contribution takes the approximate form
\begin{equation}
\frac{|a_{\rm Moon}-a_\oplus|}{a_N} \simeq \beta_0^2 \left(\frac{\Delta R_\oplus}{R_\oplus}\right)^2 <\beta_0^2\,10^{-14},
\label{EP_cham_constraint}
\end{equation}
where the characteristic values $\Phi_\oplus\sim10^{-9}$ and $\Phi_{\rm Moon}\sim10^{-11}$ have been used. Eq. \eqref{EP_cham_constraint} shows that a viable interaction must be sufficiently weak and/or sufficiently screened in the local environment.  \\

This constraint motivates a slowly varying interaction function. In the regime $\varphi\ll M_p$, a simple linear form,
\begin{equation}
f(\varphi)=f_1+f_0\varphi~~\Longleftrightarrow~~\beta(\varphi) = f_0M_p-\sigma \equiv \beta_0.
\label{beta_linear_cham}
\end{equation}
Thus the effective coupling becomes approximately constant and the screened solutions retain the standard chameleon form, with the additional Brans-Dicke contribution entering only through $\sigma$. \medskip

The linear form is further motivated by considering more rapidly varying functions. For example, if $f(\varphi)\propto f_1+f_0\varphi^n$ with $n\geq3$, then in the regime $\varphi\ll M_p$ the matter source is suppressed by higher powers of $\varphi$, substantially weakening its role in the effective potential. Similarly, an exponential interaction of the form $f(\varphi)\propto f_0\exp(f_1\varphi/M_p)$ reduces, over the same small-field regime, to an approximately constant effective coupling, $\beta(\varphi)\simeq\beta_1$. The linear choice therefore represents the simplest non-trivial interaction that preserves the scalar-matter source while remaining close to the standard chameleon construction. \medskip0

Moreover, for this construction, the assumption $\beta_0\sim\mathcal{O}(1)$ together with Eq. \eqref{beta_linear_cham} gives
\begin{equation}
\omega_{\rm BD} \sim \frac{32\pi}{\left(f_0M_p-1\right)^2} - \frac32 .
\label{omega_BD_constraint}
\end{equation}
The coefficient $f_0$ necessarily has dimensions of $M_p^{-1}$. Eqs. \eqref{coupling_constraint} and \eqref{omega_BD_constraint} demonstrate that the local-gravity constraints do not independently fix the matter-coupling coefficient and the Brans-Dicke parameter; rather, they constrain the combination that determines the effective scalar coupling $\beta_0$. 

\subsubsection{Cosmological Reconstruction}
This provides the physical motivation for adopting a slowly varying, linear scalar-matter interaction in the cosmological analysis below, where we use the equivalent Jordan-frame form
\begin{equation}
f(\phi)=\kappa(\phi+\phi_0),
\label{f_linear}
\end{equation}
where $\kappa$ determines the strength of the scalar-matter coupling and $\phi_0$ is a constant. We make an additional note here concerning the trace of the matter energy-momentum tensor. For a perfect fluid, $T^{(m)} = \rho_m-3p_m = \rho_m(1-3w_m)$, where $w_m=p_m/\rho_m$. As a result, the trace contribution to Eq.~\eqref{phi-eq} can vanish for a radiation-like fluid with $w_m=1/3$. Radiation is therefore a limiting case in which the trace-mediated part of the scalar-matter interaction disappears. However, for the present choice $L_m=-\rho_m$ and a non-constant $f(\phi)$, the term proportional to $f'(\phi)\rho_m$ in Eq.~\eqref{phi-FRW} remains nonzero. Hence, radiation is not completely decoupled from the geometric scalar. The scalar-matter interaction becomes particularly relevant for the late-time non-relativistic sector, where $p_m \simeq 0$ and $T^{(m)} \simeq \rho_m$.  \\

\begin{figure}
\begin{center}
    \includegraphics[width=0.4\textwidth]{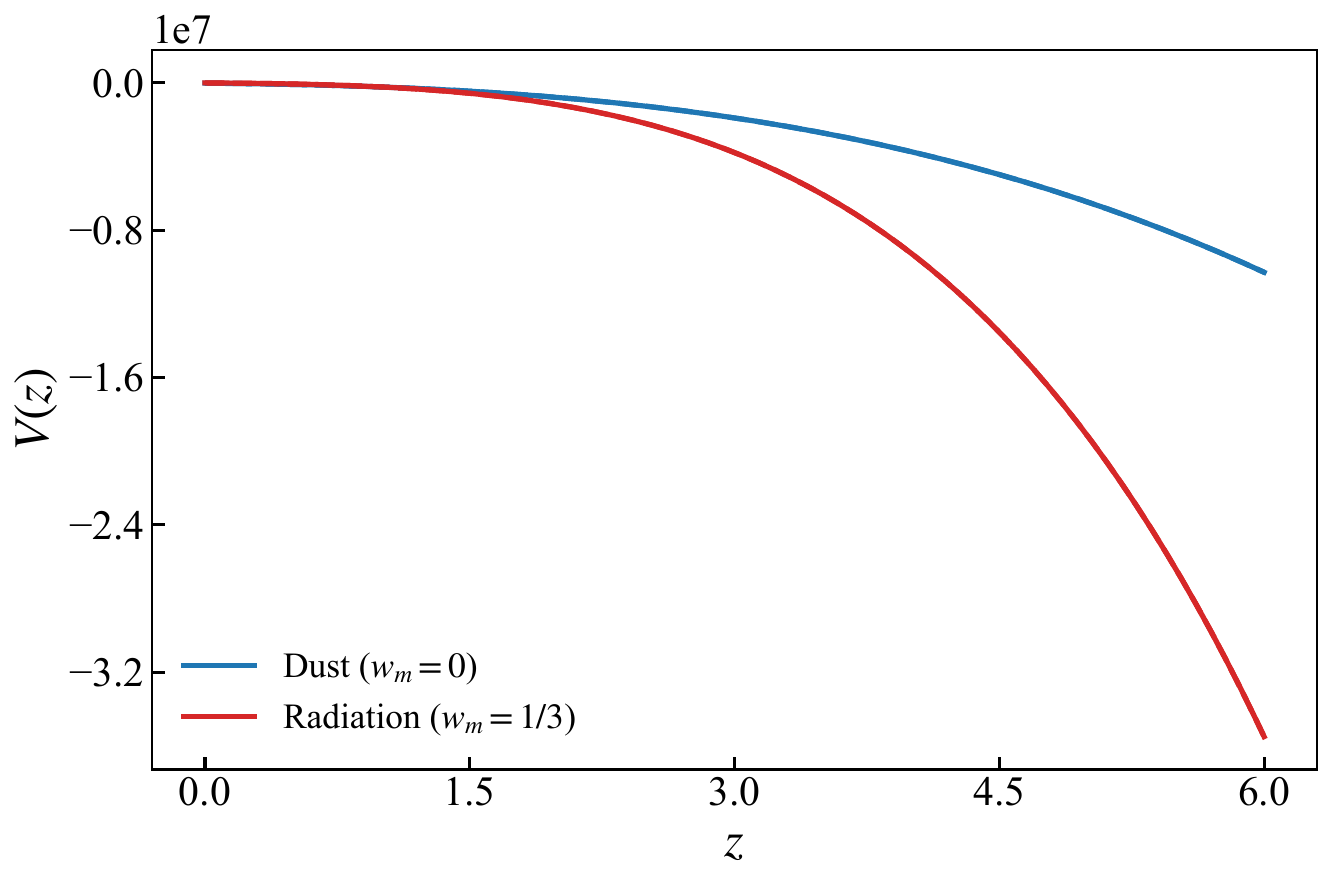}
    \includegraphics[width=0.4\textwidth]{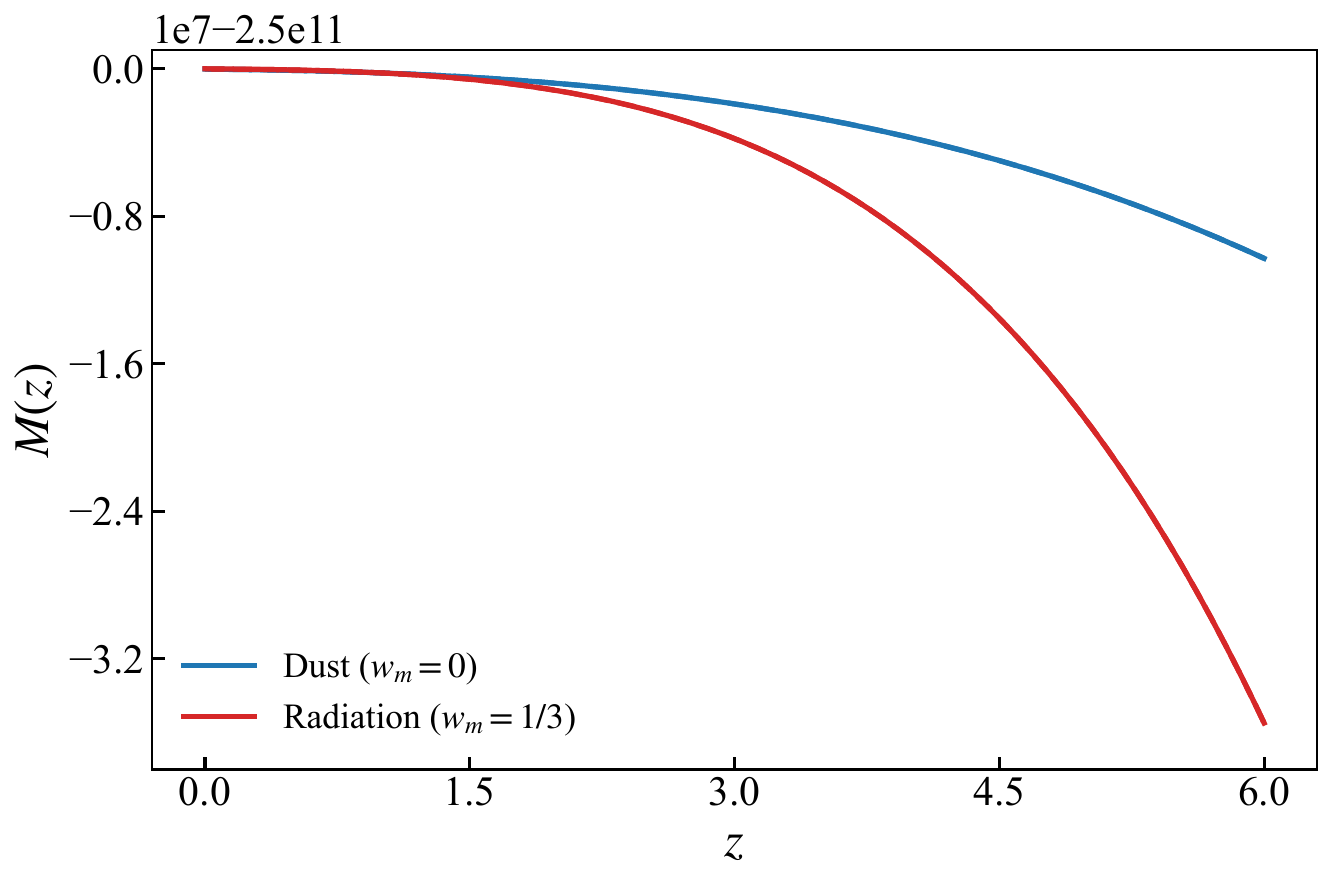}
    \includegraphics[width=0.4\textwidth]{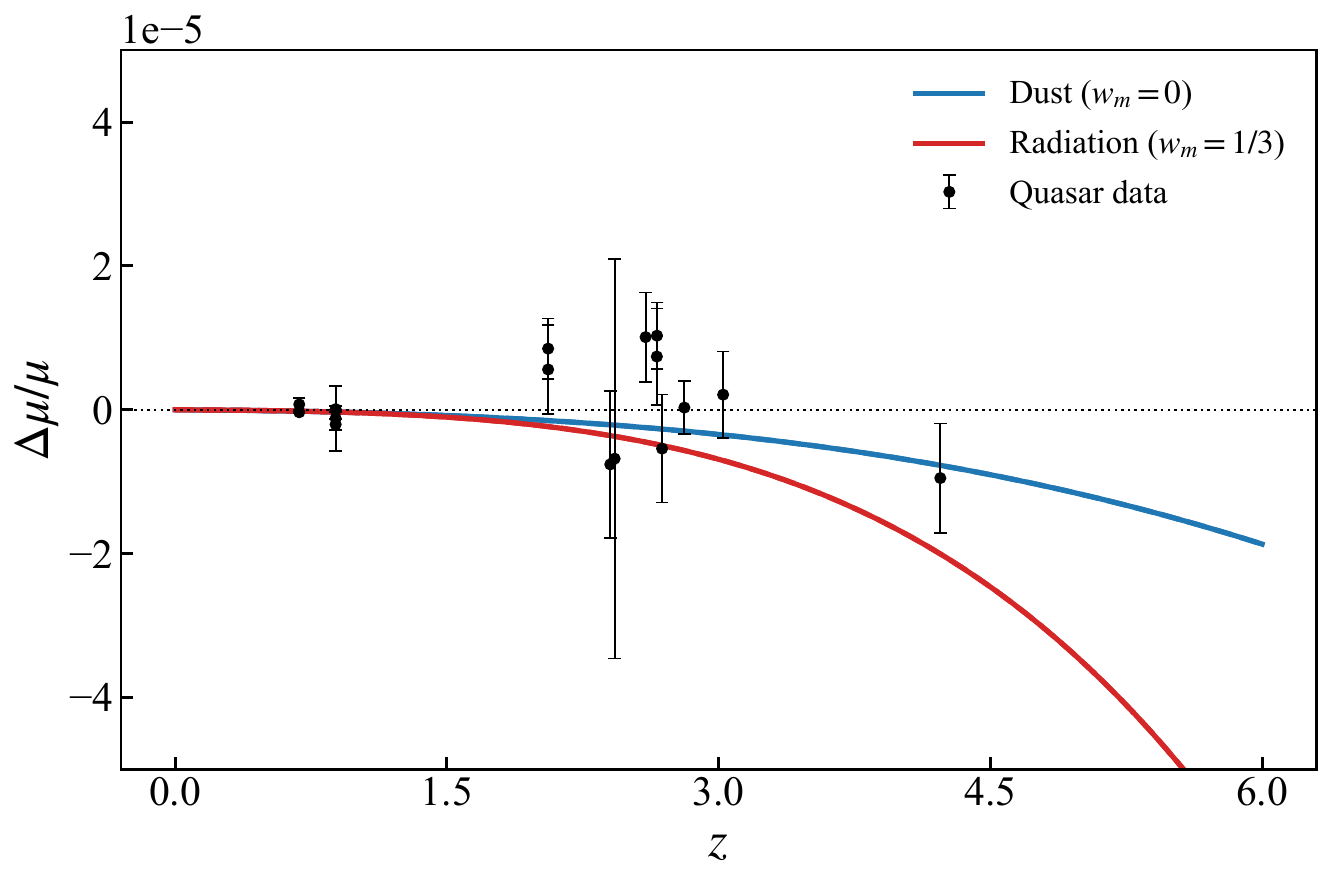}
    \caption{Reconstructed scalar potential $V(z)$, evolution of $M(z)$ with redshift and the variation of the proton-to-electron mass ratio $\Delta\mu/\mu$, with redshift in the chameleon Brans--Dicke framework for dust and radiation cases respectively.}
    \label{fig2}
\end{center} 
\end{figure}

\begin{figure}
\begin{center}
    \includegraphics[width=0.4\textwidth]{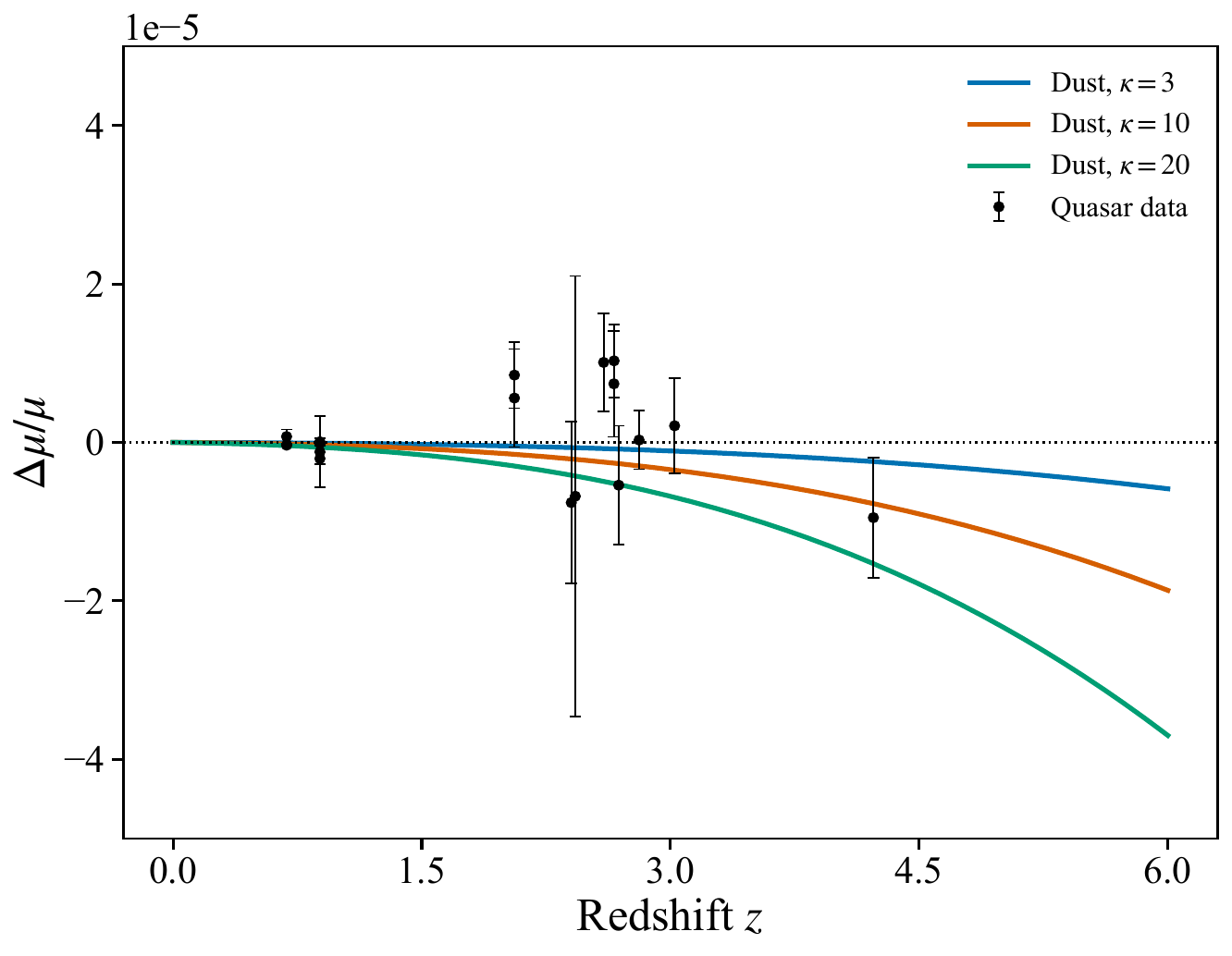}
    \includegraphics[width=0.4\textwidth]{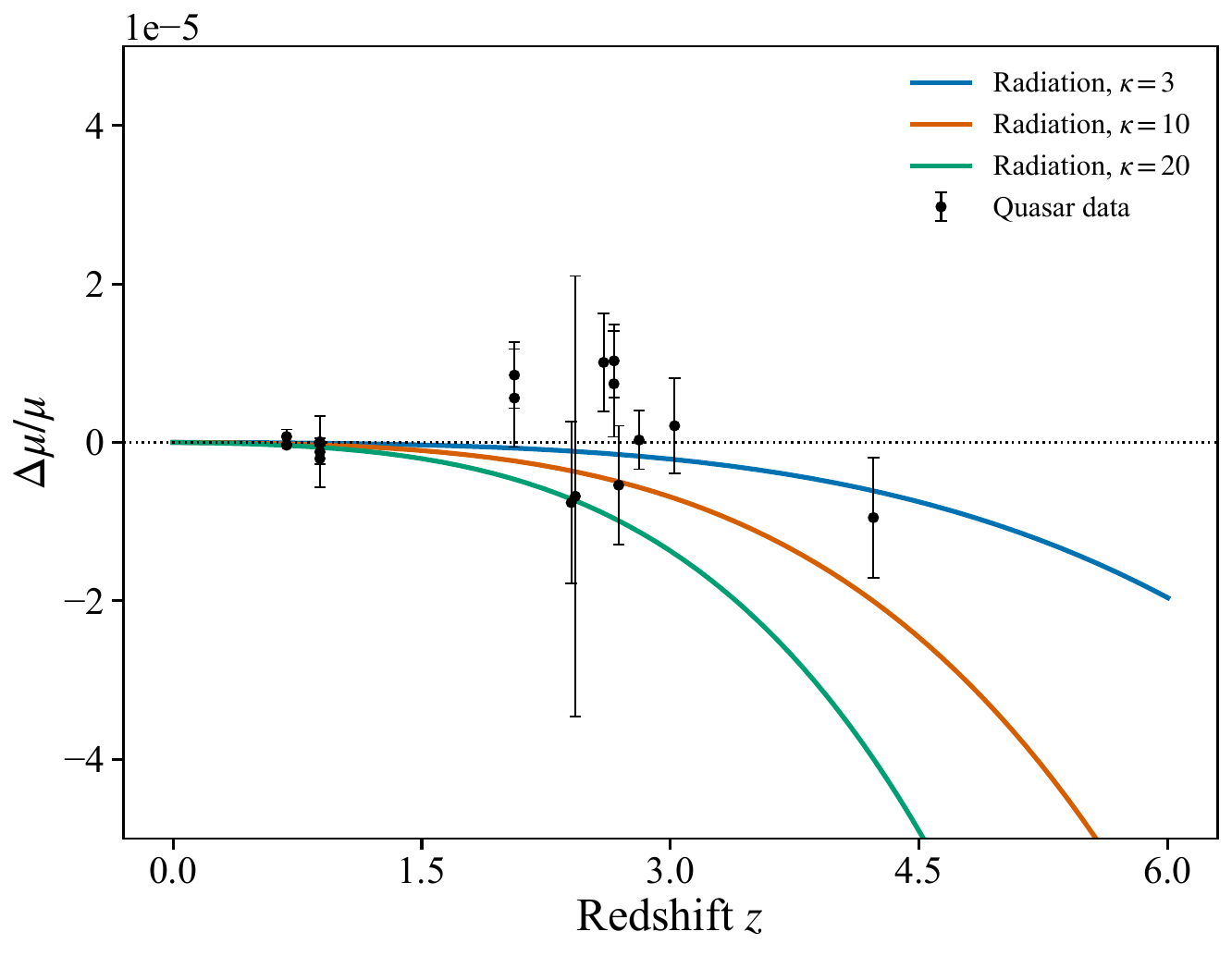}
    \caption{Evolution of $\Delta\mu/\mu$ for different $\kappa$ values for radiation and dust cases.}
    \label{fig22}
\end{center}
\end{figure}

We now substitute the coupling function $f(\phi)$ into Eqs.~\eqref{fried1}-\eqref{Vpsi} and solve the resulting system numerically for the redshift evolution of $\psi(z)$, the reconstructed potential $V(z)$, and the effective mass coefficient $M(z)$. As in the minimally coupled case, the background expansion is supplied by the observationally constrained Pad\'e-inspired form as in Eq.~\eqref{Hpade}, with the $\Lambda$CDM expansion recovered for $r_1=r_2=s_1=s_2=0$. The reconstructed profiles are then used in Eqs. \eqref{delmu} and \eqref{delmumu} to derive the evolution of $\Delta\mu/\mu$. We plot the evolution of $V(z)$ and $M(z)$ in Fig.~\ref{fig2} and $\Delta\mu/\mu$ in Fig. \ref{fig22} together with the observational constraints listed in Table~\ref{tab:deltamu}. It is evident that the chameleon-like interaction suppresses the scale of redshift variation of $M(z)$ and $\mu$, which remains well within the observational bounds.  \\

\begin{figure}
    \centering
    \includegraphics[width=0.8\linewidth]{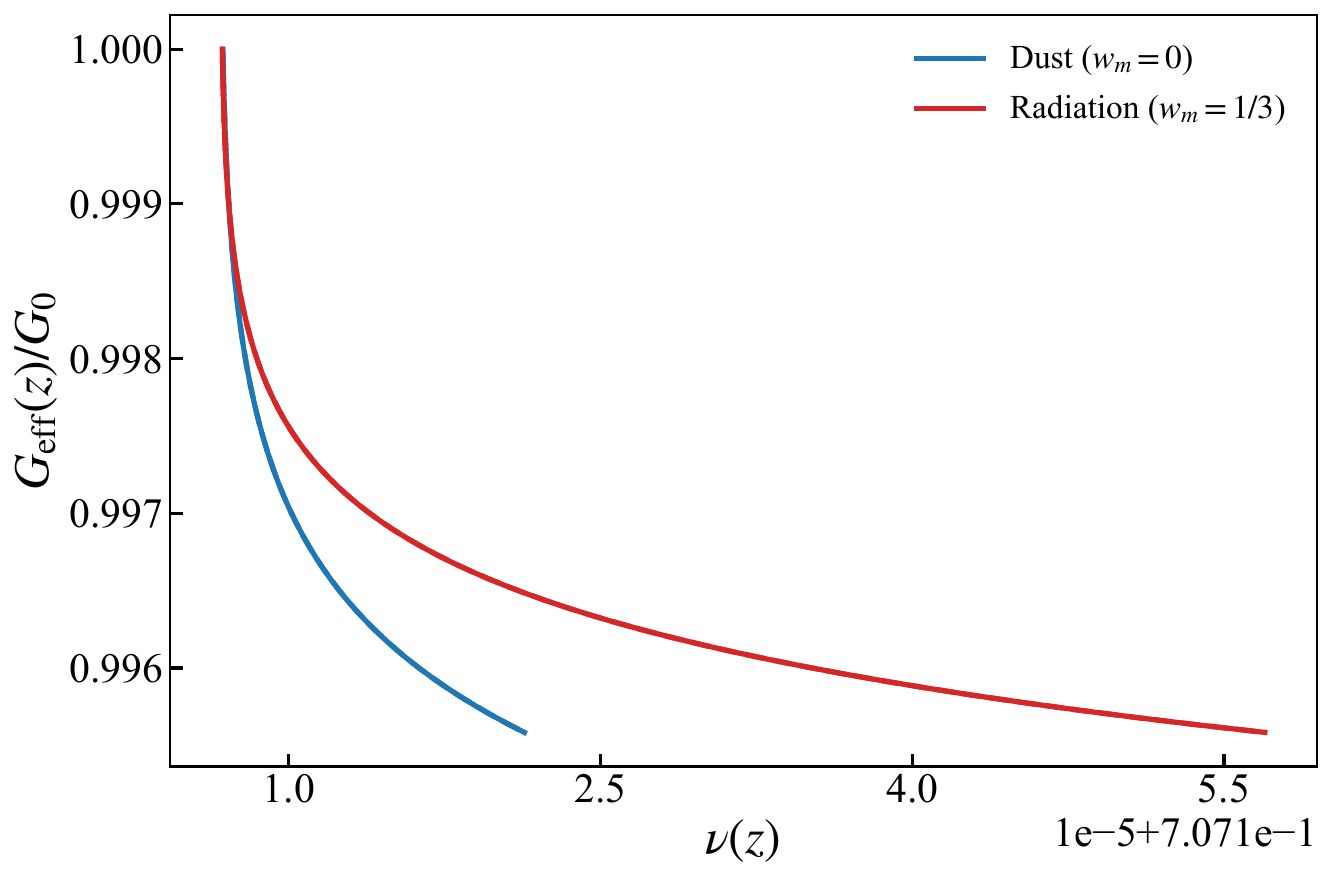}
    \caption{Evolution of $G_{\rm eff}$ as a function of the Higgs vacuum expectation value in the Chameleon Brans--Dicke framework.}
    \label{fig222}
\end{figure}

In Fig. \ref{fig222} we show a parametric evolution of the effective gravitational coupling $G_{\rm eff}$ as a function of the Higgs vacuum expectation value $\nu(z)$, within the chameleon BD framework. The behavior is not directly suggestive of any underlying mechanism, however, it is suggestive of a structure similar to Dirac's Large Numbers Hypothesis \citep{Dirac}: the gravitational and particle-physics scales may not be fundamentally independent but dynamically connected through cosmic acceleration. Unlike the original Large Numbers Hypothesis, the variations remain strongly suppressed due to the chameleon screening, which is crucial to keep consistency with current observational constraints. Our results support a possibility that scalar field driven cosmic acceleration can accommodate a common origin for a variation of gravitational coupling and particle masses.

\begin{table}
\centering
\caption{Model parameters used in the numerical analysis.}
\label{tab:model_parameters}
\begin{tabular}{l c}
\hline
\hline
Parameter & Value \\
\hline

$\phi_0$ & $1$ \\

$\nu_G$ & $0.001$ \\

$\omega$ & $1000$ \\

$H_0$ & $70$ \\

$\Omega_m$ & $0.3$ \\

$\omega_{m,\mathrm{eff}}$ & $0.3$ \\

$r_1$ & $-0.184$ \\

$r_2$ & $1.134$ \\

$s_1$ & $-0.238$ \\

$s_2$ & $0.698$ \\

$\kappa$ & $10$ \\

$\phi_{0,\mathrm{\textit{f}}}$ & $1$ \\

$\lambda$ & $10^{12}$ \\

$V_0$ & $1$ \\

$\epsilon$ & $10^{-5}$ \\

\hline
\hline
\end{tabular}
\end{table}

\section{Conclusions}
\label{sec:conclusions}

In this work, we have investigated the possibility that the cosmological evolution of the effective gravitational coupling and that of particle-mass scales may be related through a common scalar-tensor framework. The central idea is to construct a reconstruction chain connecting the late-time cosmological expansion history to a geometric Brans-Dicke scalar and its
interaction with ordinary matter. The connection allowed us to constrain the cosmological evolution of the proton-to-electron mass ratio $\mu=m_p/m_e$. \\

A first ingredient of the reconstruction is the RG-motivated logarithmic parameterization of the effective gravitational coupling, which, within the Brans-Dicke lagrangian determines the redshift evolution of the geometric scalar field. This
provides a direct phenomenological connection between the cosmological Hubble scale and the evolution of the gravitational sector. The logarithmic running is employed here as an RG-motivated phenomenological ansatz rather than as a derivation from a specific ultraviolet completion of gravity. Its usefulness in the present analysis is that it supplies a well-defined mapping $H(z)\rightarrow G_{\rm eff}(z)\rightarrow\phi(z)$, which can subsequently be propagated into the matter-sector reconstruction.   \\

For the cosmological background, we adopted a rational Pad\'e deformation of the $\Lambda$CDM Hubble function. The $\Lambda$CDM model is recovered smoothly when all Pad\'e deformation coefficients vanish, making it a well-defined limiting case of the reconstruction. A combined analysis of cosmic chronometers, BAO, DESI DR2, Type-Ia supernovae and CMB distance
information gives $H_0=73.22\pm0.26\ {\rm km\,s^{-1}\,Mpc^{-1}}$ with the remaining Pad\'e parameters allowing only modest departures from the reference $\Lambda$CDM expansion over the redshift range considered. The model-comparison statistics provide an important qualification to this result : although the Pad\'e model improves the maximum-likelihood fit by
$\Delta\chi^2=-7.453$, the corresponding $\Delta{\rm AIC}=0.547$ indicates essentially comparable statistical support, whereas $\Delta{\rm BIC}=22.503$ favours the simpler $\Lambda$CDM description. Thus, the Pad\'e parameterization should be regarded as a toy model, a controlled reconstruction framework which allows possible departures from $\Lambda$CDM. In particular, the inferred value of $H_0$ should not by itself be interpreted as evidence for a resolution of the current $H_0$ tension.   \\

Using the reconstructed expansion history, we determined the dynamics of the matter-sector scalar $\psi$. Its effective potential has an effective Higgs-like form $V(\psi,z)=V_0+M(z)\psi^2+\frac{\lambda}{4}\psi^4$, where the quadratic coefficient $M(z)$ is reconstructed dynamically rather than prescribed a priori. Whenever $M(z)<0$, the scalar possesses the symmetry-breaking minima. The evolution of $M(z)$ therefore determines the evolution of the symmetry-breaking scale. Under a Yukawa and QCD-motivated approximation, this evolution leads to the variation of the proton-to-electron mass ratio $\frac{\Delta\mu}{\mu}$. An important feature of the reconstruction is that the scalar potential can be obtained parametrically as $V(\psi)$, rather than being restricted to its original representation as a function of redshift. \\

The comparison between the two generalized Brans-Dicke frameworks is one of the central results of the present analysis. In the first case, the geometric Brans-Dicke scalar does not interact directly with ordinary matter, while the Higgs-like matter-sector scalar remains minimally coupled. In this case, the reconstructed evolution of the vacuum expectation value leads to a variation of $\mu$ that is substantially larger compared to the observationally allowed range inferred from quasar absorption spectroscopy. Thus, a minimally coupled two-scalar reconstruction of this type does not provide a satisfactory description of the observed variation of $\mu$. We then introduced a chameleon-like interaction between the geometric scalar and ordinary baryonic matter, given by $f(\phi)=\kappa(\phi+\phi_0)$, while keeping the matter-sector scalar $\psi$ minimally coupled. This additional interaction modifies the equation governing $\phi$ through matter-dependent source terms, and the altered evolution of $\phi$ feeds back into the reconstructed solution for vacuum expectation value. Consequently, the chameleon extension suppresses the evolution of the matter-sector mass coefficient and its associated vacuum expectation value. The predicted $\Delta\mu/\mu$ is reduced to the range probed by the available quasar absorption measurements. This comparison demonstrates that the scalar-matter interaction can have observable consequences beyond the usual discussion of fifth-force suppression: it can also control the cosmological evolution of an effective particle-mass scale.   \\

The main implication of the present work is the following : an RG-motivated running of the effective gravitational coupling, when embedded in a chameleon-Brans-Dicke theory, naturally leads to a testable chain of cosmological reconstruction. Additionally, the chameleon, i.e, the non-minimal interaction between the Brans-Dicke scalar and baryonic matter modifies the Higgs-like scalar sector enough to satisfy the current spectroscopic measurements of particle mass. However, we emphasize that some aspects of the present construction remain model dependent. In particular, the results depend on the assumed logarithmic form of $G_{\rm eff}(H)$, the Higgs-like form of the effective matter potential, the choice of matter Lagrangian in the non-minimally coupled theory and the functional form of $f(\phi)$. Therefore, our analysis should only be viewed as a proof of viability of the reconstruction scheme for a representative coupling rather than as a unique determination of the scalar-matter interaction. A more general exploration of the coupling functions, including their parameter degeneracies with the Brans-Dicke sector, would be a natural extension to consider.

\section*{Data Availability Statement}
The work presented in this manuscript is primarily mathematical. The data used for supplementary analysis are available on reasonable request. No new data generation is applicable.

\section*{Acknowledgements}
The authors acknowledge Vellore Institute of Technology for the financial support through its Seed Grant (No. SG20230027), year $2023$.

\bibliographystyle{mnras}
\bibliography{bibliography}
\end{document}